\documentclass{article}

\usepackage{graphicx}
\usepackage{amsmath}
\usepackage{amssymb}
\usepackage{float}
\usepackage{placeins}
\usepackage{pdfpages}

\definecolor{visprocessgray}{RGB}{127,127,127}

\definecolor{RGcolor}{RGB}{49,121,198}

\definecolor{newsleakunitcolor}{RGB}{144,144,144}

\definecolor{frontbackendcolor}{RGB}{80,159,198}

\definecolor{dataTypeColor}{RGB}{150,98,208}

\definecolor{inputoutputcolor}{RGB}{22,166,3}

\definecolor{apicolor}{RGB}{148,23,81}

\definecolor{stepcolor}{RGB}{47,82,143}

\definecolor{lisaRed}{RGB}{237,28,35}

\definecolor{lisaBlue}{RGB}{62,70,200}

\definecolor{lisaBlue2}{RGB}{1,162,232}

\definecolor{lisaYellow}{RGB}{255,201,14}

\definecolor{lisaGreen}{RGB}{33,177,76}

\definecolor{colorbrewerorange}{RGB}{253,140,90}
\definecolor{colorbreweryellow}{RGB}{255,192,1}
\definecolor{colorbrewergreen}{RGB}{146,207,94}

\definecolor{C1}{RGB}{231,69,51}
\definecolor{C2}{RGB}{26,198,83}
\definecolor{C3}{RGB}{56,110,165}
\definecolor{C4}{RGB}{11,84,1}
\definecolor{C5}{RGB}{253,148,7}
\definecolor{C6}{RGB}{0,0,109}
\definecolor{C7}{RGB}{230,28,121}
\definecolor{C8}{RGB}{118,13,21}

\definecolor{darkYellow}{RGB}{255,210,0}

\definecolor{bc1}{HTML}{66C2A5}
\definecolor{bc2}{HTML}{FC8D62}
\definecolor{bc3}{HTML}{8DA0CB}

\definecolor{transcriptblack}{RGB}{0,0,0}

\definecolor{grayText}{RGB}{102,102,102}

\definecolor{factorOne}{RGB}{255,47,146}
\definecolor{factorTwo}{RGB}{3,176,240}

\definecolor{darkblue}{HTML}{005CBB}
\definecolor{darkgreen}{HTML}{307F25}
\definecolorseries{gradient}{rgb}{last}{black!3}{black!50}
\resetcolorseries[100]{gradient}

\usepackage{color}
\usepackage{booktabs}
\usepackage{calc}
\usepackage{colortbl}
\usepackage{comment}
\usepackage{ifthen}

\newcommand{\mybarecce}[1]{%
	\textcolor{black}{\rule{0.25pt+1pt*\real{#1}}{1.5ex}}\hfill
    \ifthenelse{\equal{#1}{0}}{\textcolor{black}{{\scriptsize #1}}}{\textcolor{black}{{\footnotesize #1}}}
}

\definecolor{cRanking}{HTML}{065700}

\newcolumntype{a}{>{\columncolor{black!10}}p{6.4em}}
\newcolumntype{b}{>{\columncolor{black!0}}p{6.4em}}

\definecolor{orange}{RGB}{255,127,0}

\definecolor{blue}{RGB}{0,128,255}

\definecolor{lilac}{RGB}{158,188,218}

\definecolor{myRed}{RGB}{177,36,24}

\definecolor{myLightBlue}{RGB}{75,174,234}

\definecolor{myLilac}{RGB}{191,162,209}

\definecolor{myGreen}{RGB}{76,173,91}

\newcommand{\replicatingExaminer}[1]{\emph{replicating examiner}\,}
\newcommand{\originalExaminer}[1]{\emph{original examiner}\,}
\newcommand{\originalExperiment}[1]{\emph{original experiment}\,}
\newcommand{\replicatingExperiment}[1]{\emph{replicating experiment}\,}

\definecolor{greenSC}{RGB}{0,176,80}

\definecolor{lilac2}{RGB}{104,52,154}

\definecolor{blue2}{RGB}{56,84,146}

\newcommand{\sampleselector}[1]{\textsc{{\large S}ample{\Large S}elector}}
\newcommand{\iv}[1]{independent variable}
\newcommand{\gv}[1]{grouping variable}
\newcommand{\cv}[1]{confounding variable}
\newcommand{\con}[1]{constraint}
\newcommand{\gss}[1]{goal sample size}
\newcommand{\es}[1]{experiment sample}
\newcommand{\evs}[1]{experiment variables}
\newcommand{\dpdh}[1]{data-property-driven hypothesis}
\newcommand{\erps}[1]{experiment-relevant properties}

\definecolor{orangeAlena}{RGB}{237,125,49}

\usepackage{amsmath}
\usepackage{amssymb}
\usepackage{color}
\usepackage[normalem]{ulem}
\usepackage{booktabs}
\usepackage{calc}
\usepackage{overpic}
\usepackage{colortbl}
\usepackage{hyperref}
\usepackage{gensymb}

\usepackage[anythingbreaks,hyphenbreaks]{breakurl}
\usepackage{makecell}

\extrafloats{100}

\usepackage{subfig}

\usepackage{enumitem}
\definecolor{yellow_min_cirlcle}{RGB}{255,192,0}
\newcommand{\yellowMinCircle}[1]{\textcolor{yellow_min_cirlcle}{#1}}
\definecolor{blue_max_cirlcle}{RGB}{0,176,240}
\newcommand{\blueMaxCircle}[1]{\textcolor{blue_max_cirlcle}{#1}}
\definecolor{red_large_cirlcle}{RGB}{255,0,0}
\newcommand{\redLargeCircle}[1]{\textcolor{red_large_cirlcle}{#1}}

\usepackage{xcolor}
\usepackage{authblk}

\begin{document}
	
\title{A Shape Change Enhancing Hierarchical Layout for the Pairwise Comparison of Directed Acyclic Graphs}

\author[1]{Kathrin Guckes (n\'{e}e Ballweg)}{Kathrin.Ballweg@gris.tu-darmstadt.de}
\author[1]{Marc Sch\"{a}pers}{m.schaepers1993@gmail.com}
\author[2]{Prof. Margit Pohl}{margit.pohl@tuwien.ac.at}
\author[4]{Prof. Andreas Kerren}{andreas.kerren@liu.se}
\author[3]{Prof. Tatiana von Landesberger}{landesberger@cs.uni-koeln.de}

\affil[1]{Graphical Interactive Systems Group, Technical University Darmstadt}

\affil[2]{Informatics, TU Wien}


\affil[3]{Visualisierung und Visual Analytics, University of Cologne}

\affil[4]{Information Visualization Group, Link\"{o}ping University}

\maketitle

\begin{abstract}

Comparing directed acyclic graphs (DAGs) is essential in various fields such as healthcare, social media, finance, biology, and marketing. DAGs often result from contagion processes over networks, including information spreading, retweet activity, disease transmission, financial crisis propagation, malware spread, and gene mutations. For instance, in disease spreading, an infected patient can transmit the disease to contacts, making it crucial to analyze and predict scenarios. Similarly, in finance, understanding the effects of saving or not saving specific banks during a crisis is vital.

Experts often need to identify small differences between DAGs, such as changes in a few nodes or edges. Even the presence or absence of a single edge can be significant. Visualization plays a crucial role in facilitating these comparisons. However, standard hierarchical layout algorithms struggle to visualize subtle changes effectively.

The typical hierarchical layout, with the root on top, is preferred due to its performance in comparison to other layouts. Nevertheless, these standard algorithms prioritize single-graph aesthetics over comparison suitability, making it challenging for users to spot changes.

To address this issue, we propose a layout that enhances shape changes in DAGs while minimizing the impact on aesthetics. Our approach involves outwardly swapping changes, altering the DAG's shape. We introduce new drawing criteria:
\begin{enumerate}
    \item Criteria for maximizing outward swaps of graph changes.
    \item Criteria for reshaping the DAG by repositioning swapped changes.
    \item Criteria for handling changes that cannot be outwardly swapped.
\end{enumerate}
Our layout builds upon a Sugiyama-like hierarchical layout and implements these criteria through two extensions. We designed it this way to maintain interchangeability and accommodate future optimizations, such as pseudo-nodes for edge crossing minimization.

In our evaluations, our layout achieves excellent results, with edge crossing aesthetics averaging around $0.8$ (on a scale of $0$ to $1$). Additionally, our layout outperforms the base implementation by an average of $60-75\%$.
\end{abstract}

\section{Introduction} 
\label{sec:shape_difference_enhancing_layout_for_pairwise_comparison_of_directed_acyclic_graphs}

\begin{figure*}[tb]
    \centering
    \includegraphics[width=0.8\textwidth]{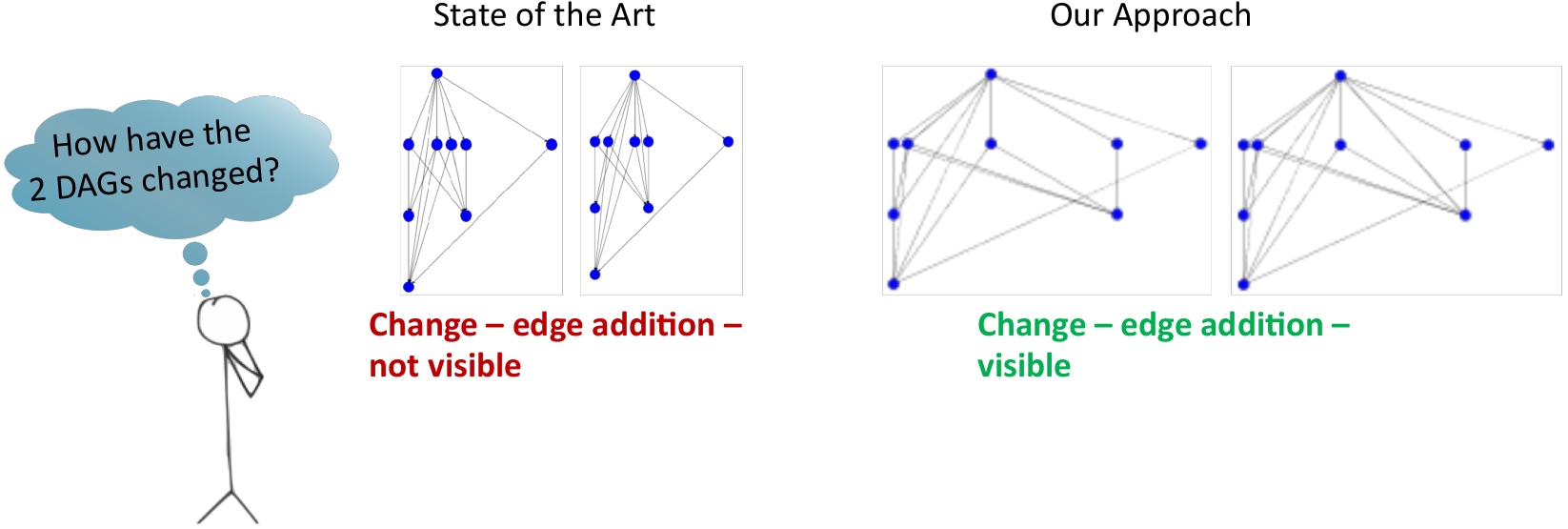}\\
  \caption[Motivational example -- shape changes of the DAGs ease their pairwise visual comparison]{Motivational example -- shape changes of the directed acyclic graphs ease their pairwise visual comparison.}
    \label{fig:teaser_CompLayout}
\end{figure*}

Visual comparison of DAGs is a task encountered in various disciplines, e.g., in healthcare, social media, finance, biology, or marketing. Directed acyclic graphs often result from so-called contagion processes over networks, that is information spreading and retweet activity~\cite{borge2013cascading,zhao2014fluxflow,archambault2016effective,viegas2013google}, disease spreading~\cite{7804480,sahneh2017gemfsim}, financial crisis propagation~\cite{von2015visual,sarlin2016macroprudential}, malware propagation in computer networks~\cite{liu2018novel}, or gene mutations~\cite{lenz2014visual}. For example, in disease spreading, an infected patient may transmit the disease to the patients, who were in contact with her. Those contact persons, in turn, may transmit the disease to their contacts along the social network. This is currently a lively topic in healthcare.  There, medical experts analyze, predict, and compare spreading scenarios ~\cite{MDRO,haarbrandt2018highmed}. In all of these applications, it is often crucial to compare two scenarios, for instance, the comparison of two disease spreading scenarios -- 1) with and 2) without patient isolation~\cite{v.20191328,haarbrandt2018highmed}. Or in the case of social media analysis, social media experts are interested in human retweeting behavior for two different message types. Necessary comparisons for financial crisis propagation analysis may be the effects of saving or not saving specific banks on the crisis contagion \cite{von2015visual}.The domain experts needs to identify even small differences between DAGs, i.e., changes of few nodes and/or edges may be crucial. The presence or absence of just on edge, as in the motivating example shown in Figure~\ref{fig:teaser_CompLayout}, can have a significant impact. For disease spreading, it may determine whether an entire hospital ward needs to be isolated or whether an entire school or entire country needs to be isolated as we experienced it with the Corona virus. Also reaching a certain recipient of a tweet may decide upon whether a tweet goes viral or not.

Visualization often supports such comparisons of DAGs. In this case, it is the visualization's job to convey the differences between the two DAGs. Differences, also called changes, may be nodes and/or edges which have been added and/or deleted. As our motivational example of Figure~\ref{fig:teaser_CompLayout} shows, even for small graphs, the visualization of changes is difficult -- especially of subtle ones like the addition of a single edge to the inner structure of the visualized DAG. As Figure~\ref{fig:teaser_CompLayout} shows, standard hierarchical layout algorithms have a troubles visualizing such changes so that they are well spottable. By standard hierarchical layout algorithms we mean layout algorithms which follow the drawing principles of the Sugiyama layout \cite{sugiyama1981methods} and respect the common graph drawing aesthetics to achieve a well-readable drawing result~\cite{battista1998graph,tamassia2013handbook,makinen2005barycenter,10.1007/978-3-7091-6215-6_19,Purchase08,Archambault11,Diehl2002,doi:10.1111/cgf.12791}. We decided for a standard hierarchical node-link diagram layout with the root placed on top (\includegraphics[height=0.4cm]{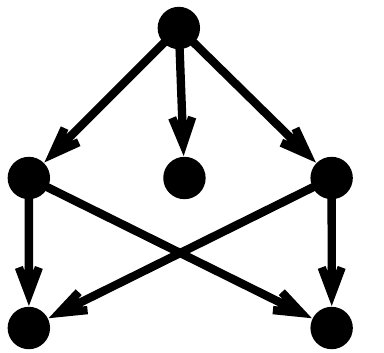}), since Burch et al. \cite{6065011,6596142} found that this layout type outperforms other types such as orthogonal or radial layouts. The problems of these standard Sugiyama-like algorithms are readily understood. The optimizations and aesthetics they consider, except for mental map stability, are not designed for comparison but for visualization of a single graph -- DAG. The mental map stability helps to prevent the position of the remaining graph elements from changing too much, thus making the comparison even more complicated for humans. However, it has no influence directly on the visualizations of the changes in the DAGs. So, this means we need optimizations that take into account comparison aspects such as good change detectability. To the best of our knowledge, such optimizations for the good detectability of changes in DAGs do not currently exist.\\
A fair question is: ``Why not just highlight the the DAG's changes?'' The answer to this questions is: it is not as easy as it seems. Scenarios where color as a visual variable is needed for other information are likely. Often graph-structured data has additional attributes like nodes or edge categories (cf. e.g.,~\cite{kerren2014network}) and color is one of the most suited visual variable for such information \cite{Bertin:1983:SG:1095597}. So, it would be really beneficial to have another visual mapping for the DAGs' changes.

In our study on visual comparisons with respect to commonalities \cite{10.1007/978-3-319-73915-1_20,JGAA-467}, we found that humans are sensitive to small variations in shape. They denoted DAGs like this \includegraphics[height=0.4cm]{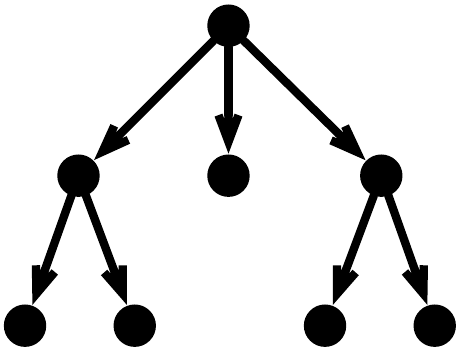} as ``narrow pyramids'' and DAGs like this \includegraphics[height=0.4cm]{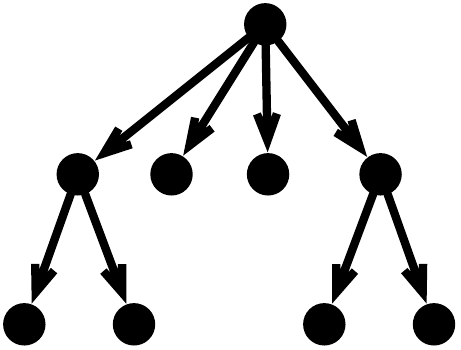} as ``wide pyramids''. So, why do we not just exploit the human sensitive notion of shape for improving the detectability of DAG changes in the context of pairwise visual comparisons? Figure~\ref{fig:pictures_DACH_ExploitationIdea} visualizes the idea: Instead of not paying particular attention to the positioning of the changing graph elements, our idea is to outwardly swap as many changes as possible, taking into account the graph drawing aesthetics, thus changing the shape of the DAG.
But this needs new drawing criteria:
\begin{itemize}
	\item We need drawing criteria which lead to outwardly swapping as many graph changes as possible.
	\item We need drawing criteria which lead to shape changes by  repositioning the outwardly swapped DAG changes.
	\item We need criteria which deal with DAG changes which are not possible to outwardly swap -- nevertheless we also need a change of the DAG's shape for those as well.
\end{itemize}

We propose a shape change enhancing layout which implements the above criteria based on two extensions of a Sugiyama-like hierarchical layout. Our layout performs the shape change enhancements so that the influence on graph aesthetics is as small as possible. We decided to implement the novel criteria extension-based to allow for the interchangeability of the Sugiyama-like layout on which we based our layout. Interchanging the base Sugiyama-like layout may be necessary in case of further optimizations -- e.g., pseudo-nodes for further edge crossing minimization -- are needed. In our case with employing the barycenter edge crossing minimization method, further edge crossing minimization was not needed since on average we achieve values of around $0.8$ for the edge crossing aesthetic. These are very good results since a value $1$ is the highest possible value. Furthermore, our evaluation shows that our layout outperforms the base implementation, on average, in $60-75\%$.


\subsection{Related Work}
\label{sc:rw}

Since we propose a shape change enhancing layout for the pairwise visual comparison of DAGs which fosters the detectability of changes for humans by changing the DAG shape, our work is located in these research areas:
\begin{itemize}
	\item Static graph drawing resp. layouts
	\item Dynamic graph drawing resp. layouts
	\item Graph layouts based on user study results
	\item Graph drawing aesthetics
\end{itemize}

Static graph drawing is related since our algorithm type of choice is rooted in the domain of static graph drawing. Pairwise visual comparison, in our case of DAGs, can be transferred into the domain of dynamic graph drawing. Let $G_1$ (base graph) and $G_2$ (alternative) be a pair of DAGs differing by $N$ node and/or edge changes. Given that per time step $t$, starting with $t_0$, $N$ changes happen then it is $(G_{1}, G_{2}) = (G_{t_0}, G_{t_{0+N}})$. This relationship can be repeated for every two graphs in the set of dynamic graphs -- i.e., pairwise visual comparison is basically a comparison of two graphs of the dynamic series which, consequently, makes dynamic graph drawing work related to ours. We based our layout development on the results of our previous commonalities- and difference-coined visual comparison studies (cf. Chapter~\ref{cha:research_goal_1_influence_factors_on_the_human_similarity_notion_of_directed_acyclic_graphs_in_the_context_of_pairwise_comparison}). Consequently, other graph layouts resulting from such an approach are related. Graph drawing aesthetics are well-accepted measures for assessing the drawing quality of a layout result. So, for sure, they also play a pivotal role for our work -- i.a. because we have to ensure that our shape changes only have little influence on the already optimized graph aesthetic criteria. 



\subsubsection{Static Graph Layouts}
    \label{sc:rwlayoutstatic}
There are various types of algorithms for laying out node-link diagrams. Amongst others, there are Sugiyama-like layouts (cf. e.g., \cite{tamassia2013handbook,battista1998graph}), orthogonal layouts (cf. e.g., \cite{orthogonalExample,burch2011visualizing}), ballon or bubble layouts (cf. e.g., \cite{Grivet2006,JGAA-153}), and radial layouts (cf. e.g., \cite{orthogonalExample,battista1998graph}). However, Burch et al. \cite{6065011,6596142} found that Sugiyama-like layouts with the root placed on top outperforms other types such as orthogonal or radial layouts when it comes to humans working with the visualization. Sugiyama-like layouts are variants of the seminal layout algorithm of Sugiyama et al.~\cite{sugiyama1981methods}.These layout variants differ mainly in the optimization algorithm used for maximizing aesthetics -- examples for edge crossing optimization are i.a. global or local sifting, median, or the barycenter method~\cite{kaufmann2003drawing,bachmaier2011global}. Barycenter and median are easy to implement and lead to better and faster results than sifting \cite{bachmaier2011global}. For DAGs, in particular, edge crossing minimization and the preservation of symmetries are relevant \cite{battista1998graph,bennett2007aesthetics}.

\subsubsection{Dynamic Graph Layouts}
    \label{sc:rwlayoutdynamic}
While specialized layouts may be developed, a widely applied strategy is to use a static layout (see above) and extend it for the dynamic case. An important feature of the dynamic layout is the preservation of the mental map~\cite{Purchase08,Archambault11}. It means a stable visualization where``the placement of existing nodes and edges should change as little as possible when a change is made to the graph.''~\cite{bachmaier2011global}: Methods for ensuring stable node-link visualizations of dynamic graphs have been extensively explored (cf. i.a.~\cite{archambault2014}). In early papers on this topic, this notion has been termed the ``preservation of the mental map''~\cite{eades:1991:C,misue1995layout}. Today, it is rather denoted as ``drawing stability''. A number of possible property optimizations to achieve mental map preservation were proposed in these works. These include orthogonal ordering, topology, scaling, horizontal shuffle, and force-scan. Most of this work, and other work emerging around the same time~\cite{paulisch:1990:SPE}, considered the scenario of a user interacting with a visualization where nodes were added/removed/re-positioned by user interactions and a balance between drawing stability and quality needed to be struck. The observance of this criterion is ensured by using so-called supergraphs for offline graph drawing. This is also applicable to our case, as we know the compared graphs in advance. In this case, a supergraph is constructed from the input graphs and then the supergraph is laid out. This can be a strict or elastic case. The strict case is also called foresighted layout without tolerance. The strict case positions all nodes at the same location across all graphs the supergraph is constructed from. It fully ensures the preservation of the mental map~\cite{Diehl2002,10.1007/978-3-7091-6215-6_19}. 

Drawing stability has been evaluated extensively in the graph drawing literature from metric-focused~\cite{10.1007/978-3-642-25878-7_11,8580419} and human-centered~\cite{Archambault11,10.1007/978-3-642-36763-2_42,Ghani:2012} experiments. From a human-centered perspective, the benefits of a stable drawing -- especially of a foresighted layout without tolerance -- is in specific node and path identification, where the human can offload cognitive effort to the visualization as she knows common parts will remain in the same place during graph evolution~\cite{Archambault16}. This is also really helpful for spotting graph changes \cite{Franconeri2014}. Therefore, we use a foresighted layout without tolerance.

\subsubsection{Graph Drawing Aesthetics}
    \label{sc:rwaesthetics}

Graph drawing aesthetics are well-accepted measures for assessing the drawing quality of a layout result. They have been mainly researched for static graph drawings. Important aesthetic criteria can be deemed to be edge crossing minimization, equal edge length, minimal node overlap, edge angle, graph symmetry, edge bending~\cite{LIN2018123,purchaseaesthetics,purchase2002metrics,Purchase:1997a,10.1007/BFb0021827,kobourov2014crossings}. These aesthetic criteria, or a subset of them, are optimized by graph layout algorithms. In our work, we build upon these works and extend them with additional drawing criteria for pairwise comparison leading to graph shape changes. We have to ensure that our shape changes only have little influence on the already optimized graph aesthetic criteria. 


\subsubsection{Graph Layouts based on User Study Results}
Huang et al. \cite{huang2018making} postulate the need for the consideration of human notions, perception, or cognitive processes for graph visualization since according to Huang et al. \cite{huang2018making} ``to produce truly user cognitively friendly visualizations, well-grounded cognitive theories and design guidelines are needed.'' Amongst others, Kieffer et al. \cite{Kieffer2016}, Coleman et al. \cite{bachmaier2011global}, Siebenhaller et al. \cite{10.1093/bib/bby099}, and Lin et al. \cite{Lin:2021aa} employed an approach where their final layout considers human notions, perception, or cognitive processes to make the graph visualization cognitively friendly for humans. 

Moreover, researchers assessed with which layout users can work best (cf. e.g., \cite{6065011,s.20201039}) or how users create graph layouts (cf. e.g.,~\cite{4658147,dwyer2009comparison}). Specifically for hierarchic data, Burch et al. \cite{6065011,6596142} and Archambault et al. \cite{archambault2016effective} analyzed the influence of graph layout on graph readability. They found that Sugiyama-like layouts are best readable for trees and DAGs.






\subsection{Shape} 
\label{sub:shape}
\begin{figure}[tb]
    \centering
    \includegraphics[width=0.4\linewidth]{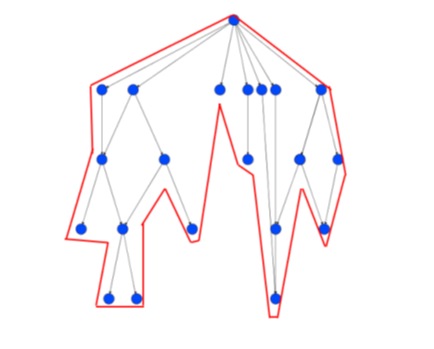}
        \includegraphics[width=0.3\linewidth]{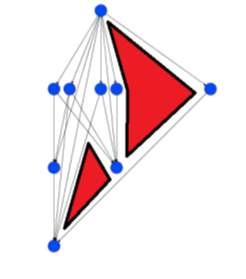}
    \caption[Shape in the graph visualization domain]{Shape in the graph visualization domain -- outer shape (left -- red line), inner shape (right -- red areas). (Figure based on original Figures from \cite{MarcMA})}
    \label{fig:shapetypes}
\end{figure}

In our study on visual comparisons with respect to commonalities (cf. Section~\ref{sec:influence_factors_for_comparisons_with_respect_to_commonalities}), we found that humans are sensitive to small variations in shape. They denoted DAGs like this \includegraphics[height=0.4cm]{pictures/DACH/smallPyramid.eps} as ``narrow pyramids'' and DAGs like this \includegraphics[height=0.4cm]{pictures/DACH/largePyramid.eps} as ``wide pyramids''. Further recent studies have found the same across different visualization types: scatterplots~\cite{10.1145/2858036.2858155}, star plots~\cite{doi:10.3138/carto.44.3.217,Klippel:2009}, star glyphs~\cite{Fuchs:2014}, matrix order \cite{7534849}, and geometric shapes for cartographic representations \cite{ijgi8050217}. Herewith, we find strong substantiation of our idea to exploit the human sensitive notion of shape for improving the detectability of DAG changes in the context of pairwise visual comparisons.

As we found with our two visual comparison studies, in the context of pairwise visual comparison the factor shape consists of the outer and the inner shape of the DAG. What exactly does this mean? We would like to explain this using the doughnut and Bavarian doughnut analogy: If we asked you what the shape of a doughnut is, you would answer ``A circle with a hole in the middle -- something like this \includegraphics[height=0.3cm]{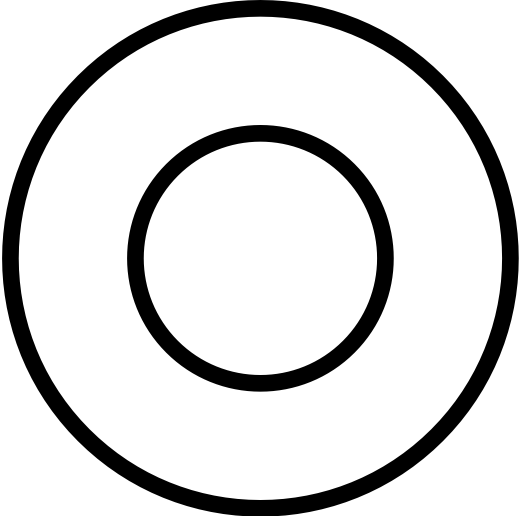}.''. If we asked you what the shape of a Bavarian doughnut is, you would answer ``A circle -- something like this \includegraphics[height=0.3cm]{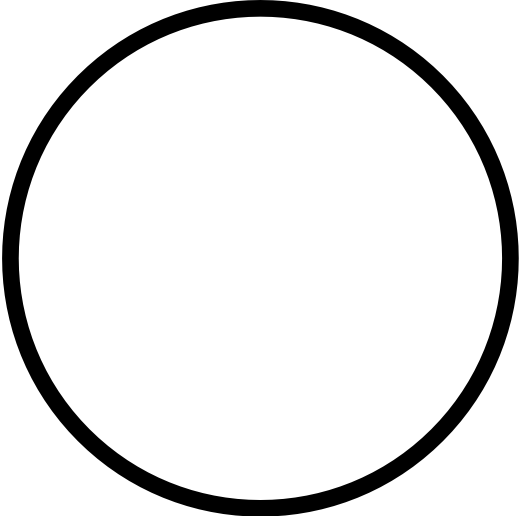}.'' The ``circle'' is the outer shape and the ``hole in the middle'' is the inner shape. Transported into the domain of graph visualization -- the outer shape is a hull which encloses the visualized node-link diagram and the inner shape are the white space areas which result from laying out the DAGs and visualizing them as node-link diagrams (cf. Figure~\ref{fig:shapetypes}).
\FloatBarrier

\subsubsection{Outer Shape -- Hull} 
\label{ssub:outer_shape_hull}

\begin{figure}[tb]
    \centering
    \includegraphics[width=0.95\linewidth]{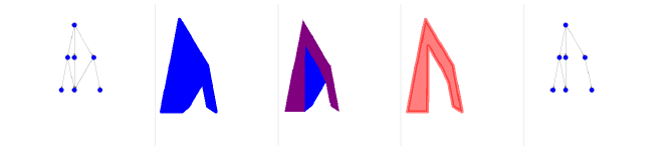}\\
    \includegraphics[width=0.95\linewidth]{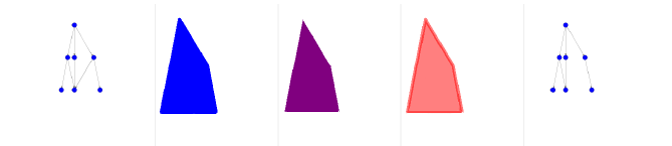}
       \caption[Example based illustration of the influence of the hull type on the shape and also the shape change of DAGs visualized as node-link diagrams]{Example based illustration of the influence of the hull type on the shape and also the shape change of DAGs visualized as node-link diagrams.}
    \label{fig:hullchanges}
\end{figure}

A shape of visual objects is often defined using a so-called hull, contour, boundary, outline, or enclosing polyline. All are different names for the same concept of defining a visual object covering, i.e., encompassing or including, visual elements of an object~\cite{schreck2008butterfly,1532142,Fuchs:2014}. There is a large variety of shapes for objects in 2D, in particular objects consisting of a finite set of 2D points~\cite{edelsbrunner2010alpha}. Shapes can cover all elements, like the convex hull does, or only parts of the elements -- e.g., without outliers~\cite{Schreck:2007aa}. As we are interested in changes of the DAGs' hull we aim for shapes that enclose all elements.

The shape can have various forms, such as a bounding box, a bounding circle, a triangle, or more complex shapes such as a convex hull, a concave hull, a butterfly, or an alpha shape~\cite{schreck2008butterfly,Schreck:2007aa,1532142,GRAHAM1972132,1056714,edelsbrunner2010alpha}. The hull can directly bound the elements like the convex or the concave hull and the alpha shape, build so-called skeletons (e.g., MST~\cite{1532142}), or can be more distant from the elements, e.g., hulls using so-called distance fields~\cite{10.1145/1921264.1921286} or the alpha-disks -- shapes using centered circles~\cite{edelsbrunner2010alpha}. 

The shapes may include holes or not~\cite{schreck2008butterfly}. The holes refer to 'empty' local areas, where no objects are located. These areas are fully surrounded by areas with objects, In mathematical geometry topology, they also identify objects of genus-1 or above~\cite{munkres2018elements}. Shapes with holes, are more general than only outer shapes -- outlines~\cite{1056714}. An example of such a general shape is alpha-shape and its variants~\cite{edelsbrunner2010alpha}. The local empty areas correspond to the notion of the inner white space in our work.
%

Often, shape refers to the outer hull of an object such as the concave or the convex hull~\cite{10.1007/978-3-319-27261-0_41}. As shown in Figure~\ref{fig:hullchanges}, the choice of the hull can have impact on the shape change of DAGs visualized as node-link diagrams. The top image shows that adding an edge to a DAG can lead to a shape change for a concave hull, but the shape remains constant for the convex hull. In other cases, an addition of one edge can change also the convex hull. Thus, the choice of the hull type has an impact on shape change.
All above mentioned shape definitions refers to shapes of 2D points. However, graphs resp. DAGs include not only points (= nodes), but also connecting lines (= edges). Figure~\ref{fig:hullchanges} shows that edges may play an important role for shape change. Related work on graph shapes (cf. e.g.,~\cite{10.1007/978-3-319-27261-0_41}) focuses on large graphs where the graph shape can be approximated solely by points (= nodes).  Moreover, the paper by Eades et al.~\cite{10.1007/978-3-319-27261-0_41} refers to undirected general graphs. Directed acyclic graphs laied out with a Sugiyama-like layout in 2D space have also a special top-down form, that should be considered. Owing to the variety of shape definitions and the need for graph shapes with edges, it is challenging to find a suitable shape definition for DAGs. In order to address this challenge, we conducted a qualitative user study which showed the DAGs' shape type perceived by humans.

\paragraph{Qualitative User Study.} 
\label{par:qualitative_user_study}
\begin{figure}[tb]
  \centering
    \includegraphics[width=\textwidth]{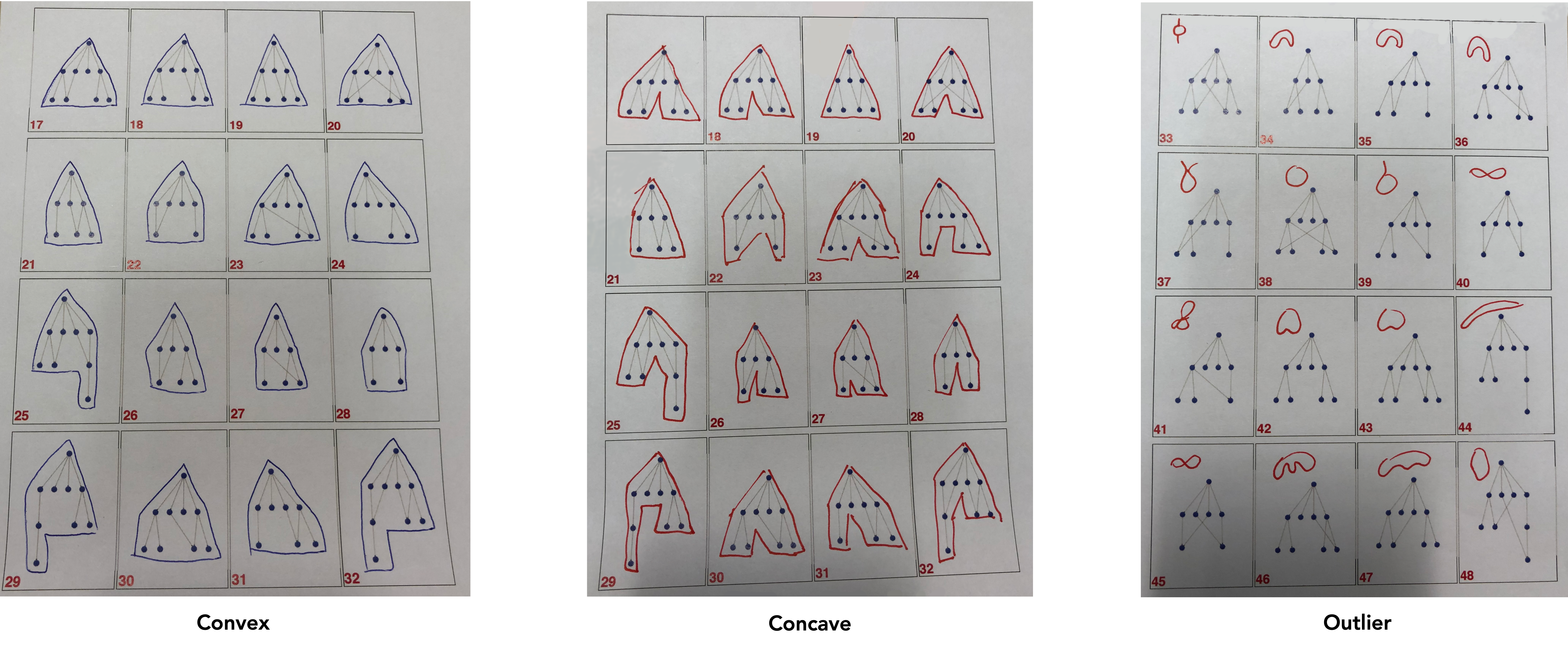}
  \caption[Qualitative user study on the shape of DAGs -- examples of the hull types our participants drew]{Qualitative user study on the shape of DAGs -- examples of the hull types our participants drew. $11$ out of $20$ participants drew a concave hull. Six participants drew convex hulls and three participants were outliers.}
  \label{fig:pictures_DACH_ShapeLayout_QualitativeUserStudy}
\end{figure}

We asked twenty participants to draw the outer shape of a set of DAGs. We used the same DAGs as we used for our visual comparison study with respect to commonalities and also presented the visualized DAGs as a print out. The participants should draw the shape of the DAGs with a pen. We neither trained nor instructed participants on this task. As for the influence factors for the human similarity notion, we wanted to find out which shape humans see when they are asked for the outer shape of a DAG visualized as a top-rooted node-link diagram without any priming which training or instructions may cause \cite{doi:10.1080/13658810701517096,doi:10.1111/cgf.13409,eurovisshort.20181082,10.1145/2556288.2557200,10.1145/2556288.2557200}.

Two coders evaluated the participants' drawings. Both coders employed a top-down qualitative content analysis~\cite{Schreier:2012,Saldana:2012}. This method is especially used for studying verbal material, but is also used for images \cite{EGBERGTHYME2013101}. It is a systematic method based on a system of categories -- a coding scheme.

The result of the coders' coding consensus was that the participants drew three types of shapes (cf. Figure~\ref{fig:pictures_DACH_ShapeLayout_QualitativeUserStudy}):
\begin{itemize}
	\item Convex hulls
	\item Concave hulls
	\item Outliers
\end{itemize}

$11$ out of $20$ participants drew a concave hull (cf. Figure~\ref{fig:pictures_DACH_ShapeLayout_QualitativeUserStudy} -- \texttt{Concave}). Six participants drew convex hulls and three participants were outliers (cf. Figure~\ref{fig:pictures_DACH_ShapeLayout_QualitativeUserStudy} -- \texttt{Convex}, \texttt{Outlier}). They, for instance, applied affine transformations to the visualized DAGs to achieve more complex formations like something which resembles the infinity symbol ($\infty$). The results show a tendency for the concave hull. So, we proceed with the concave hull. But, certainly, there is still room for detailing the result -- 1) how tightly the hull enclose the DAG and 2) when participants tend to draw a mixture of both a concave and a convex hull. To make our algorithm also usable when these results are clarified, we offer input parameters that allow to define the hull shape between convex and concave as well as how tightly the hull should enclose the DAG. For our calculations we tuned those parameters so that the characteristics of the resulting concave hull like degree of enclosure and hull shape resembles the characteristics of the concave hulls drawn by our participants(degree of enclosure: $10\%$, hull shape -- concavity: $80\%$). The calculation of the enclosure degree is in line with alpha-shape circles \cite{edelsbrunner2010alpha} and distance fields by Kalbe el al.~\cite{10.1145/1921264.1921286}.
\FloatBarrier

\paragraph{Our Outer Shape Algorithm.} 
\label{par:our_outer_shape_algorithm}

\begin{figure}[tb]
        \centering
		\subfloat[Horizontal rays which determine the left and the right \newline side of the DAG's outer shape.]{\includegraphics[width=.5\linewidth]{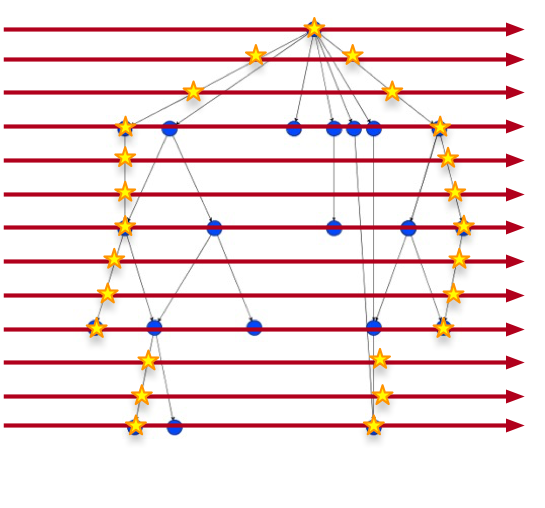}}
        \centering
		\subfloat[Vertical rays which determine the bottom and the top \newline side of the DAG's outer shape.]{\includegraphics[width=.5\linewidth]{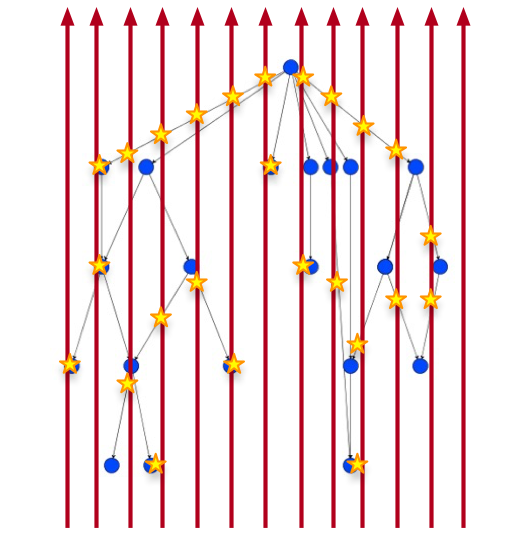}}
		\newline
        \centering
		\subfloat[Resulting concave hull.]{\includegraphics[width=.5\linewidth]{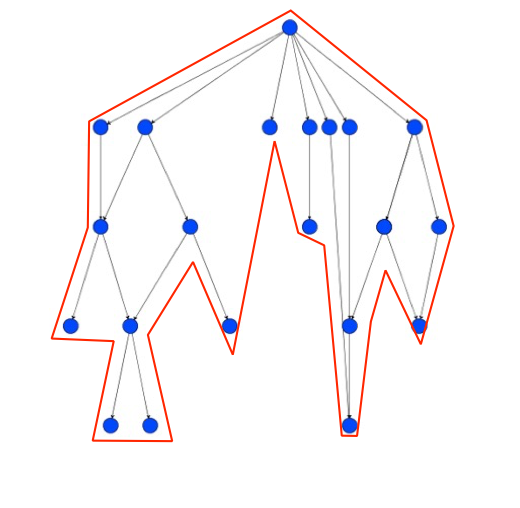}}
    \caption[Outer shape algorithm -- schematic representation]{Outer shape algorithm -- schematic representation: The intersection points of the rays and the nodes resp. edges are the points which finally form the concave hull which is shown in Figure c). (Figure based on Figures from \cite{MarcMA})}
    \label{fig:shapeRays}
\end{figure}

Convex and concave can be seen as two extremes of a continuous space of hulls enclosing all objects of the node-link diagram with a closed polyline. The concave hull 'cuts out' empty space around the enclosed objects (cf. Figure~\ref{fig:shapeRays} -- \texttt{(c)}). These hulls are commonly calculated for 2D points. The challenge was to calculate the concave hull that is suitable for Sugiyama-like, top-rooted node-link diagrams -- i.e., consider edges -- and corresponds to the shapes drawn by the study participants. We developed an algorithm that calculates the hull using a ray-tracing-like approach, which takes advantage of the NLDs' top-rootedness for sending out the rays (cf. Figure~\ref{fig:shapeRays} -- \texttt{(a)}, \texttt{(b)}). 

To be able to also consider edges, we approximate nodes by points and edges by sets of points. The number of points on the edge is configurable via the number of rays. Larger number of rays increases the precision, but leads to longer runtime. Horizontal and vertical parallel rays, cf. Figure~\ref{fig:shapeRays} a), b) -- red arrows, identify most-left, most-right, most-top and most-bottom nodes and edges. The intersection points of the rays and the nodes resp. edges are the points which finally form the concave hull (cf. Figure~\ref{fig:shapeRays} a), b) -- yellow stars; Figure~\ref{fig:shapeRays} c)). The spacing of the rays determines the number and size of the concave hull's 'cut-outs'. According to the study, the nodes positioned in unity space\footnote{nodes which are positioned according to the layouts horizontal spacing parameter -- in our case: $80=p_{HS}$} should not produce cut-outs, and larger spaces should produce cut-outs. Therefore, the spacing of the rays depends on the horizontal spacing parameter of nodes $p_{HS}$. Based on the results of our qualitative user study, we use $20=p_{HS}/4$ points. We tried up to $80$ points.

\FloatBarrier

%


\subsubsection{Inner Shape -- White Space} 
\label{ssub:inner_shape_white_space}

Inner shapes are holes –- local areas without any elements \cite{schreck2008butterfly,edelsbrunner2010alpha}. White space in node-link diagrams are local areas without nodes and without edges. They have very low local visual graph density ~\cite{dunne2015readability}. For sets of 2D points, alpha shapes, and circular hulls \cite{Schreck:2007aa,edelsbrunner2010alpha} have holes, when there are no points within a circle of a pre-set radius parameter. We extend this idea for node-link diagrams, and define white space as the area enclosed by a closed polygons of nodes and edges (cf. Figure~\ref{fig:shapetypes} -- right). 

  \vspace{-0.6cm}
\begin{figure}[tb]
	  \vspace{-0.6cm}
  \centering
  \vspace{-0.6cm}
  \subfloat[Bounding rectangle on an example DAG.]{\begin{minipage}{0.7\textwidth} \centering\includegraphics[width=\textwidth]{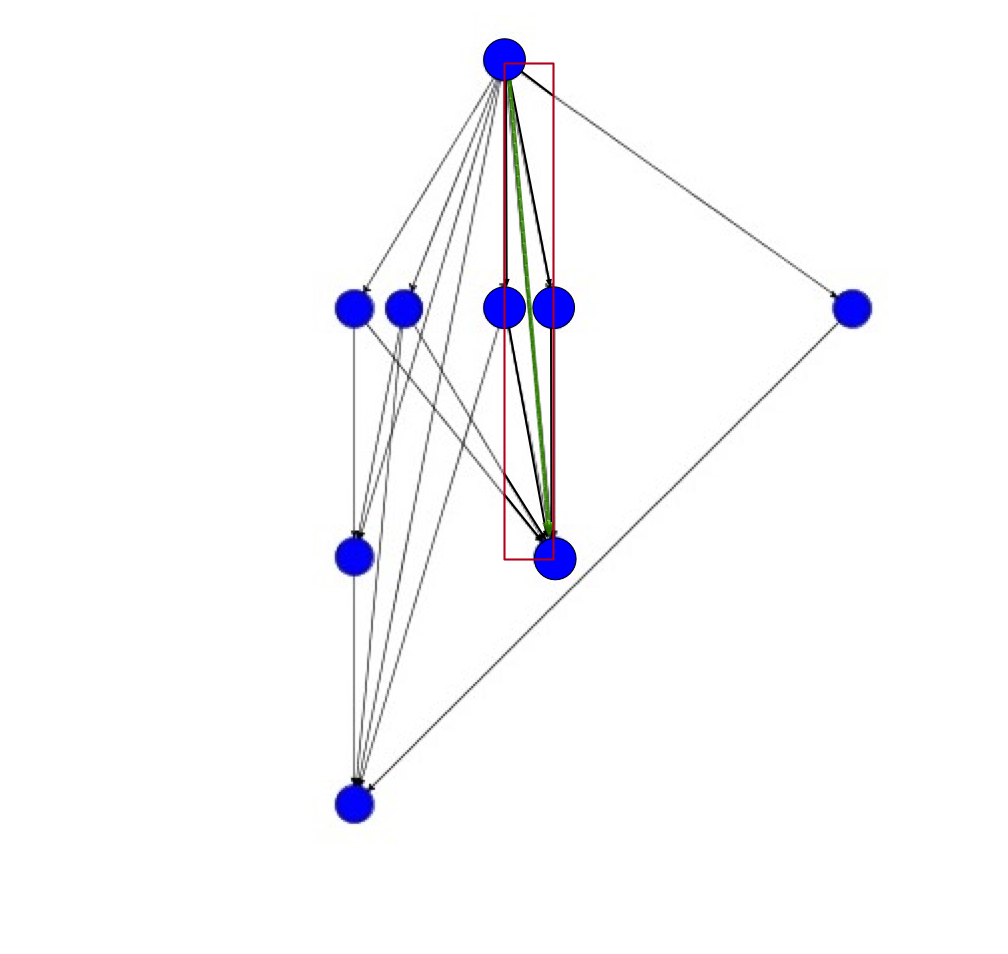}\end{minipage}}
  \newline
    \centering
  \subfloat[Subdivision of the bounding rectangle into \newline the polygons of nodes and edges.]{\begin{minipage}{0.4\textwidth} \centering \includegraphics[height=9cm]{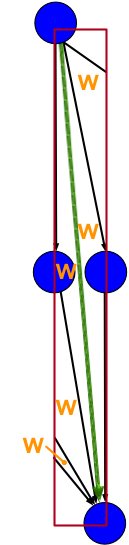}\end{minipage}}
    \centering
    \subfloat[Largest inner circles of the polygons of nodes and edges.]{\begin{minipage}{0.4\textwidth} \centering \includegraphics[height=9cm]{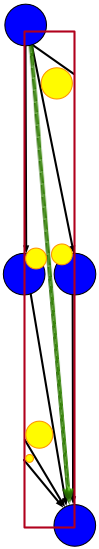}\end{minipage}}
  \caption[Inner shape algorithm -- schematic representation.]{Inner shape algorithm -- schematic representation. (Figure based on Figures from \cite{MarcMA})}
  \label{fig:pictures_DACH_ShapeLayout_WhiteSpaceCaclulation_RectangleEnclosesChange}
\end{figure}

\paragraph{Our Inner Shape Algorithm.} 
\label{par:our_inner_shape_algorithm}
Our white space algorithm works within the framework of a bounding rectangle (cf. Figure~\ref{fig:pictures_DACH_ShapeLayout_WhiteSpaceCaclulation_RectangleEnclosesChange} a) -- \texttt{red rectangle}). Within this, our algorithm searches for all faces enclosed by polygons of nodes and edges. Initially, a rectangle polygon is defined that corresponds to the bounding rectangle. Now the initial polygon is divided into new polygons by each edge that intersects the rectangle (cf. Figure~\ref{fig:pictures_DACH_ShapeLayout_WhiteSpaceCaclulation_RectangleEnclosesChange} b)). This is done until all edges that intersect the rectangle have been accounted for. For all these created polygons we calculate the largest inner circle and estimate the white space area with the circular area of the largest inner circle (cf. Figure~\ref{fig:pictures_DACH_ShapeLayout_WhiteSpaceCaclulation_RectangleEnclosesChange} c)). Finding the center point for an arbitrary polygon is not a trivial problem. For runtime reasons, we use the algorithm of Agafonkin et al. for this calculation \cite{polylabel}. This is an extension of the algorithm of Garcia-Castellanos et al. \cite{doi:10.1080/14702540801897809}.

If the bounding rectangle encloses a graph change, as in Figure~\ref{fig:pictures_DACH_ShapeLayout_WhiteSpaceCaclulation_RectangleEnclosesChange}, then the width of the bounding rectangle for edges is defined by the coordinates of the nodes that this edge connects. For a node change, the width of the bounding box is defined by the nodes' positions adjacent to the changing node and remaining the same.

\FloatBarrier









\subsection{Our Shape Change Enhancing Hierarchical Layout Algorithm} 
\label{sub:our_shape_change_enhancing_hierarchical_layout_algorithm}

\begin{figure}[tb]
  \centering
    \includegraphics[width=.9\textwidth]{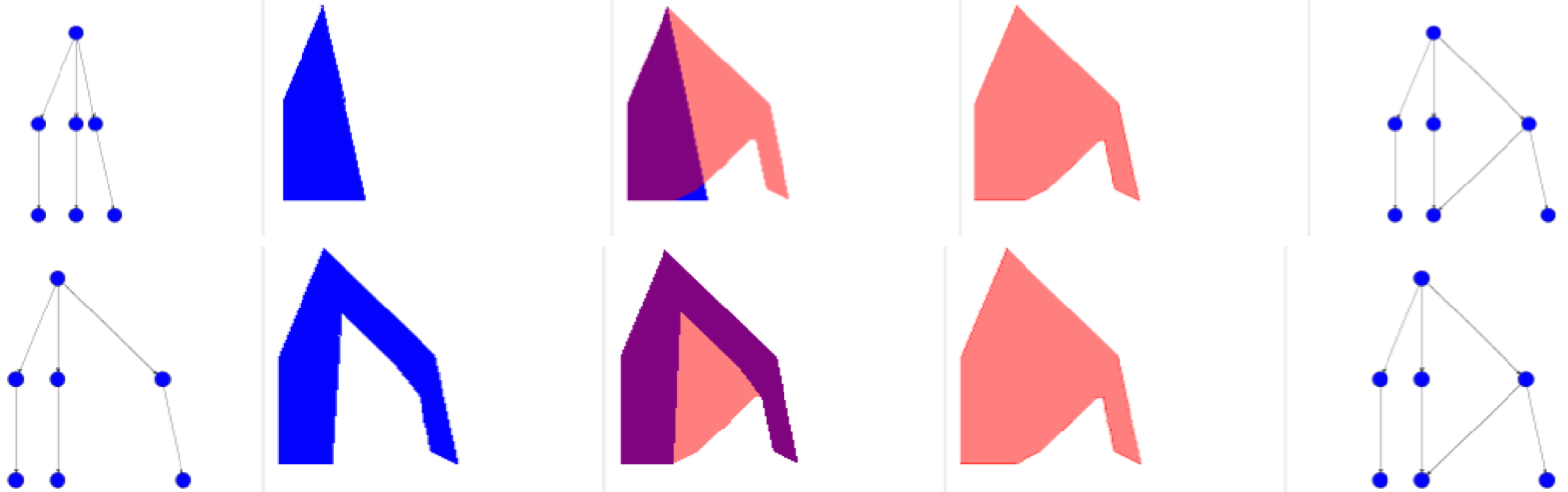}
  \caption[Shape change enhancing hierarchical layout algorithm -- effects of (no) drawing stability tolerance on the outer hull]{Shape change enhancing hierarchical layout algorithm -- effects of (no) drawing stability tolerance on the outer hull: The top row shows the approach with tolerance and the bottom row shows the approach with no tolerance. In the top row, we can clearly see that the approach leads to a large shape change. However, it also leads to a change of shape in areas of the outer hull where no graph change happened. We believe that this is why the approach with tolerance leads to shape changes which are likely to suggest the human viewer that the graph change is larger than the change that actually happened. In the bottom row, we can see that the no tolerance approach also leads to a considerable shape change. This shape change is in the area of the outer hull where the actual graph change is located.}
  \label{fig:pictures_DACH_ShapeLayout_TotalDrawingStability}
\end{figure}

Our shape change enhancing layout is based on a Sugiyama-like hierarchical layout from the JUNG graph drawing Library \cite{JUNG}. We chose a Sugiyama-like layout since Burch et al. \cite{6065011,6596142} found that this layout type outperforms other types such as orthogonal or radial layouts. As the layer assignment algorithm, our base implementation uses the longest path layering. The benefits of this layering algorithm are that it is linear with respect to runtime complexity and it produces DAG layouts with the minimum number of layers \cite{tamassia2013handbook}. This avoids the hierarchical layouts to become long and thin which negatively impacts graph readability \cite{tamassia2013handbook}. Further, it uses the barycenter method for edge crossing minimization. According to E. Mäkinen et al. \cite{makinen2005barycenter}, the barycenter method is a very good solution regarding runtime complexity and result: ``The time complexities of all of these new methods are clearly bigger than that of the barycenter heuristic. Since the number of edge crossings decreased is only marginal, we prefer the barycenter heuristic.'' It also optimizes the graph drawing's visual symmetry and employs a foresighted approach with no tolerance. An increased visual symmetry improves the readability of a graph drawing \cite{10.1007/978-3-319-27261-0_50,10.1007/BFb0021827}. Purchase et al. \cite{purchase2002metrics} even consider visual symmetry as one of the most important graph drawing aesthetics. According to Archambault et al. \cite{Archambault12}, a strict drawing stability is in general important for the viewer's orientation in the graph. In addition to that, we found with a further user study evaluating different layout approach ideas that an approach with tolerance creates shape changes which are likely to suggest the human viewer that the graph change is larger than the change that actually happened. Figure~\ref{fig:pictures_DACH_ShapeLayout_TotalDrawingStability} shows an example. The top row shows the approach with tolerance and the bottom row shows the approach with no tolerance. In the top row, we can clearly see that the approach leads to a large shape change. However, it also leads to a change of shape in areas of the outer hull where no graph change happened. We believe that this is why the approach with tolerance leads to shape changes which are likely to suggest the human viewer that the graph change is larger than the change that actually happened. In the bottom row, we can see that the no tolerance approach also leads to a considerable shape change. This shape change is in the area of the outer hull where the actual graph change is located.
%
%
%

\begin{figure}[tb]
  \centering
    \includegraphics[height=20cm]{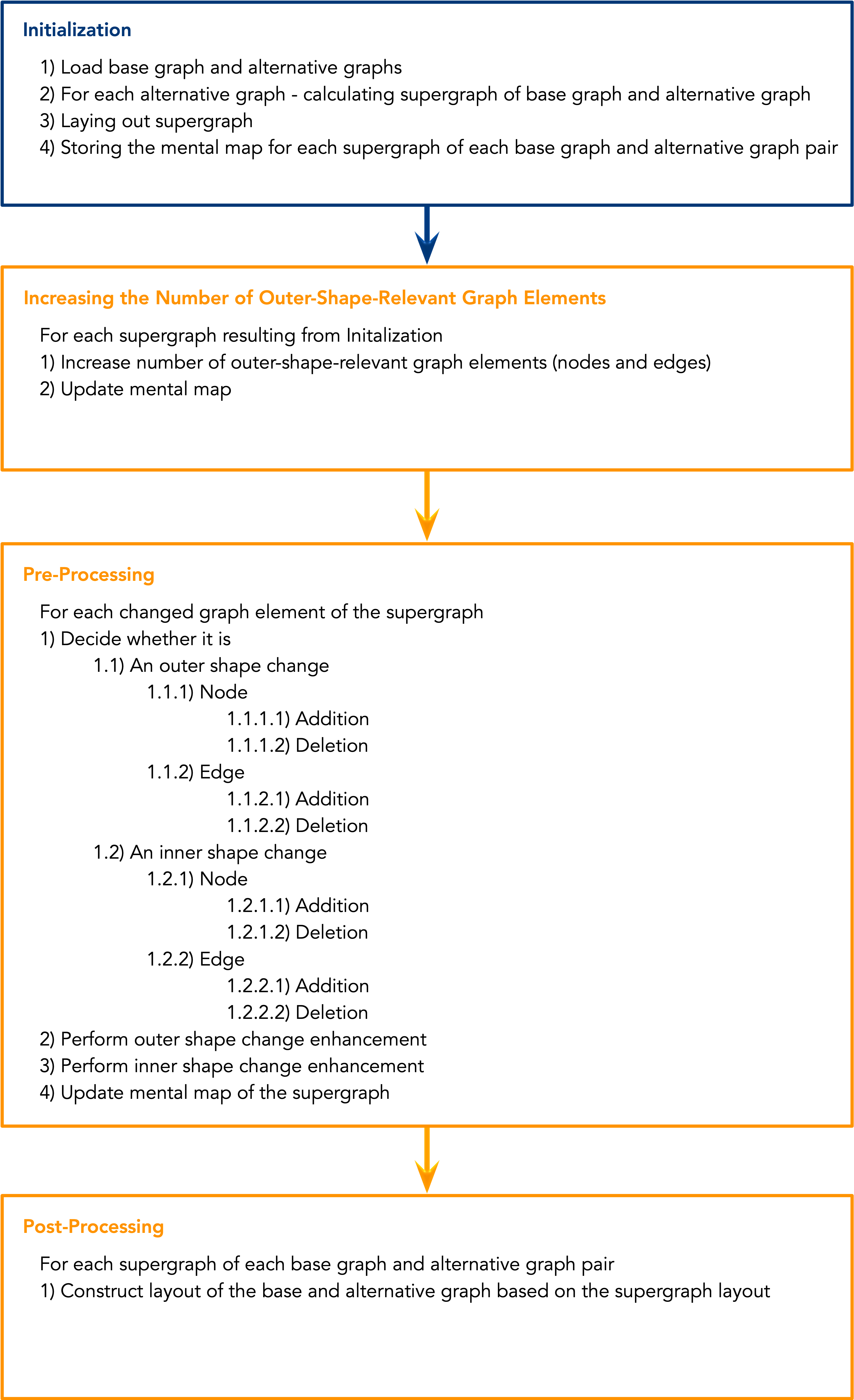}
  \caption[Shape change enhancing layout -- schema]{Shape change enhancing layout -- schema.}
  \label{fig:pictures_DACH_ShapeLayout_ShapeLayout_Schema}
\end{figure}

Our shape change enhancing layout extends this base implementation with a phase for increasing the number of outer-shape-relevant changes and a pre- and a post-processing phase (cf. Figure~\ref{fig:pictures_DACH_ShapeLayout_ShapeLayout_Schema}). Herewith, our layout implements the novel drawing criteria needed for enhancing the shape change:
\begin{itemize}
	\item We need drawing criteria which lead to outwardly swapping as many graph changes as possible.
	\item We need drawing criteria which lead to shape changes by  repositioning the outwardly swapped DAG changes.
	\item We need criteria which deal with DAG changes which are not possible to outwardly swap -- nevertheless we also need a change of the DAG's shape for those as well.
\end{itemize}

The extension-based principle was already successfully used by Diehl and his colleagues \cite{10.1007/978-3-7091-6215-6_19} to visually prepare the graph layout for changes that happened over time. This principle has two main advantages:
\begin{itemize}
	\item it ensures that the new drawing criteria are based on the state of the art criteria and layout result since the extension happen after the Initialization phase.
	\item it allows to change the state of the art Sugiyama-like layout implementation from the Initialization phase since the extension happen after the Initialization phase.
\end{itemize}

\subsubsection{Initialization} 
\label{ssub:initalization}

Our layout requires at least one one base graph and at least one alternative graph which shall be visually pairwise compared to the base graph. Throughout the initialization, we calculate for each base graph and alternative pair the supergraph. We lay out each supergraph with the afore explained base Sugiyama-like hierarchical layout and store the mental map. Laying out the supergraph with our base implementation ensures that the graph drawing is optimized with respect to the state of the art graph drawing aesthetics like minimized edge crossings or optimized visual symmetry. The following steps work under the premiss to keep their effect on the optimized aesthetics as low as possible.

\subsubsection{Increasing the Number of Outer Shape Relevant Graph Elements} 
\label{ssub:increasing_the_number_of_shape_relevant_graph_elements}

\begin{figure}[tb]
  \centering
    \includegraphics[width=.9\textwidth]{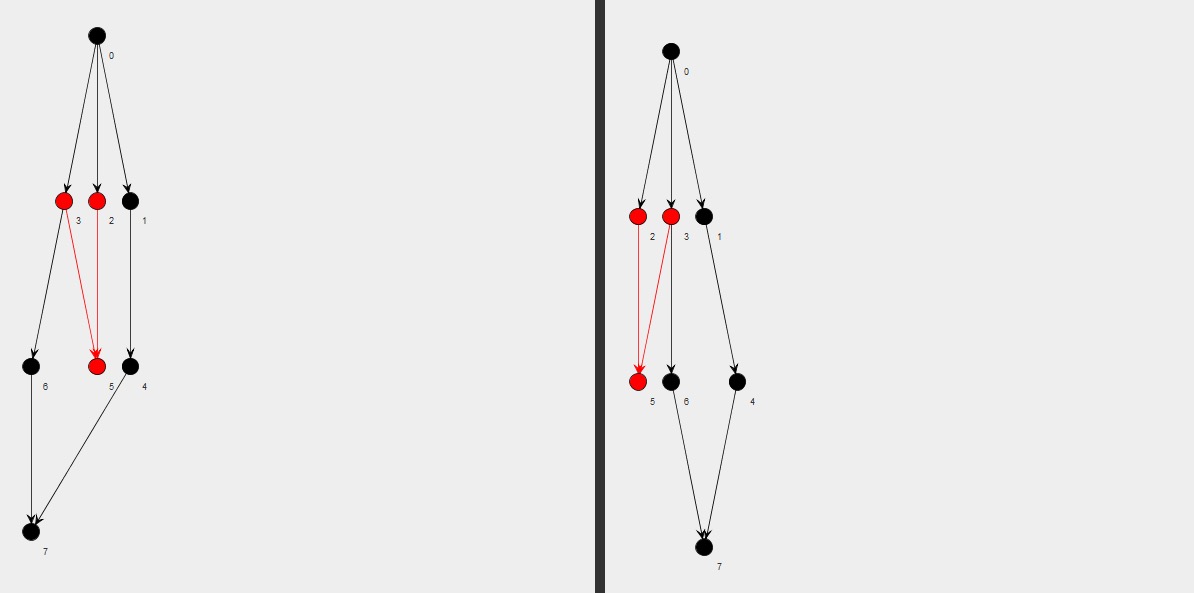}
  \caption[Increasing the number of outer shape relevant graph elements -- algorithm: example algorithm result]{Increasing the number of outer shape relevant graph elements -- algorithm: example of the detected subgraph which is shifted to the outer left of the example DAG. (Figure taken from \cite{MarcMA})}
  \label{fig:pictures_DACH_ShapeLayout_Number_of_Shape_relevant_Elements}
\end{figure}

The graph is examined for leaves that are not yet hull-relevant. Subsequently, a heuristic is started that builds a node chain recursively to the first graph level. The subgraph is repositioned to the left as well as to the right of the graph, as long as the aesthetics criteria are not degraded too much. Figure~\ref{fig:pictures_DACH_ShapeLayout_Number_of_Shape_relevant_Elements} shows a successful attempt to increase the number of hull-relevant elements.

In detail this means: First, all leaves in the graph that are not hull-relevant are identified. They are potential anchor points for the movement of a subgraph. The extension iterates the list of leaf nodes that are not hull-relevant at this point. This list must be updated after each attempt to increase the number of hull-relevant elements, since leaf nodes may also have become hull-relevant due to a move action. The method starts at the first node of the non-hull-relevant leaf node list. First, a backup of the current positions is created. This backup helps to restore the initial state should the move action negatively influence the optimized graph aesthetics criteria too much.\\
Recursively, the parent nodes are collected up to the first level -- level below the root node. Only the first parent node is taken always. As soon as the level below the root is reached, the recursion ends. At the level below the root, all parent nodes are included in the collection list. We do this to avoid straight chains of nodes and edges. If it happens that all nodes of the first DAG level are included in the collection list, either the leftmost or the rightmost node of the first level is omitted from the collection list for the further procedure, depending on the currently tested subgraph movement direction. The reason for this and for adding only one parent node in the lower levels is that otherwise there would be no more graph elements defining whether the subgraph to be moved is to the right or to the left of the actual graph.\\
All collected nodes are first moved to the right of the graph. Then a new layout process starts to update the mental map and absolute positions. As long as the aesthetics criteria are not negatively affected too much, the solution is considered acceptable. The node is thus removed from the list of non-hull-relevant leaf nodes. If the solution is not acceptable -- i.e. the aesthetics criteria are too strongly affected -- the layout is reset to the old state and the same procedure is tested on the left side of the graph. If there is also no acceptable solution on the left side, the layout is also reset.  The node is thus removed from the list of non-hull-relevant leaf nodes. The method ends when all potential nodes have been handled.
\FloatBarrier

\subsubsection{Pre-Processing} 
\label{ssub:pre_processing}
Here, we explain the idea behind the algorithm and show the idea and algorithm results based on Figures. 

\begin{figure}[tb]
  \centering
    \includegraphics[width=.7\textwidth]{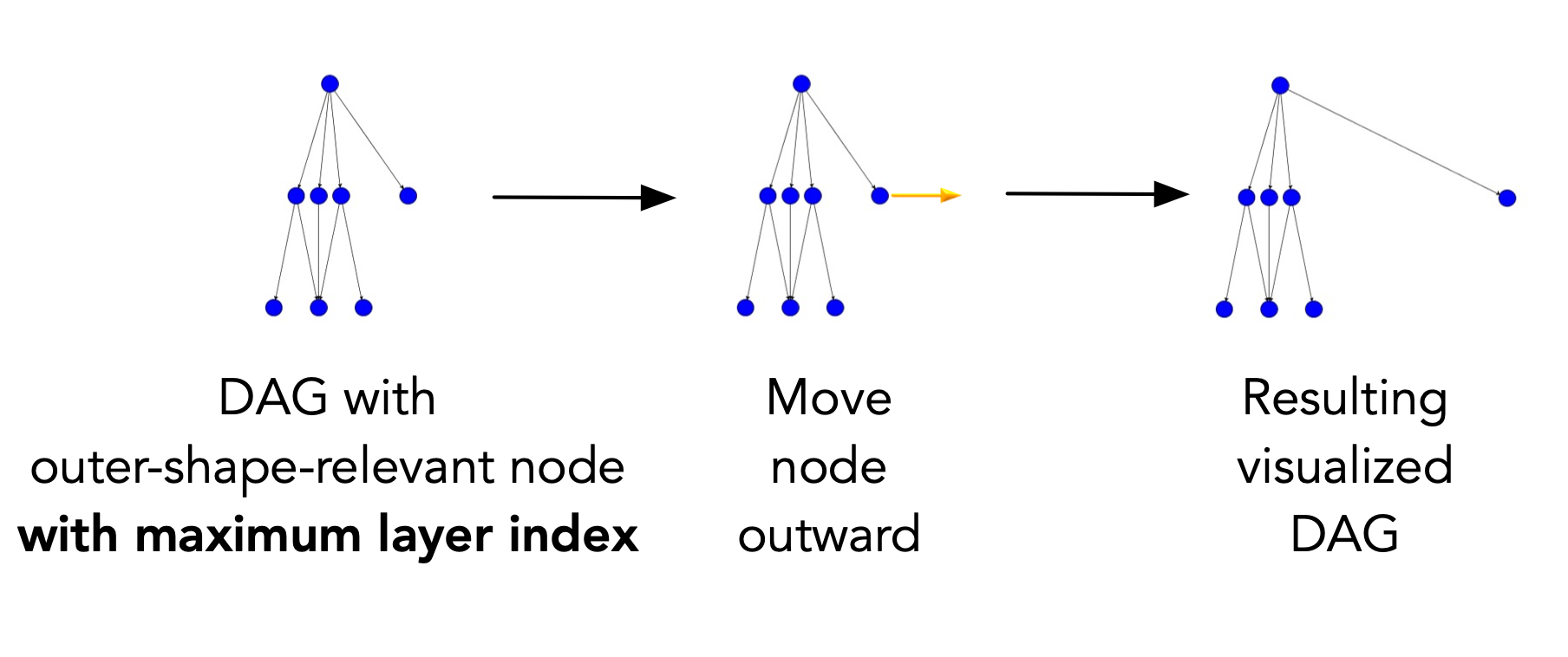}
  \caption[Outer shape change enhancement -- outer-shape-relevant node changes \& node has minimum or maximum layer index: schematic representation]{Outer shape change enhancement -- outer-shape-relevant node changes \& node has minimum or maximum layer index: schematic representation.}
  \label{fig:pictures_DACH_ShapeLayout_OuterShapeEnhancement_outerNode}
\end{figure}

\begin{figure}[tb]
  \centering
    \includegraphics[width=.7\textwidth]{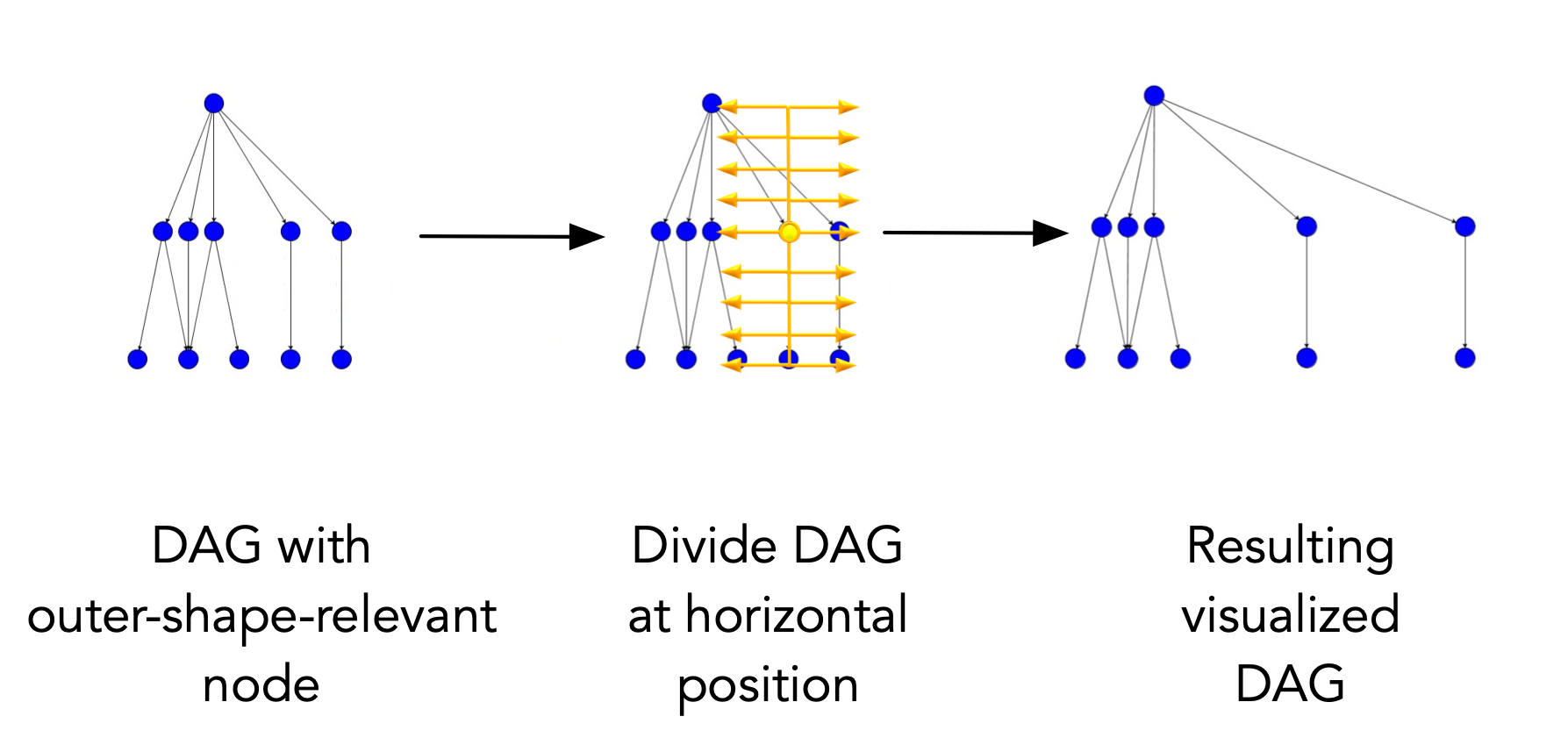}
  \caption[Outer shape change enhancement -- outer-shape-relevant node changes \& node does not have minimum or maximum layer index: schematic representation]{Outer shape change enhancement -- outer-shape-relevant node changes \& node does not have minimum or maximum layer index: schematic representation.}
  \label{fig:pictures_DACH_ShapeLayout_OuterShapeEnhancement_outerInnerNode}
\end{figure}

\begin{figure}[tb]
  \centering
    \includegraphics[width=.7\textwidth]{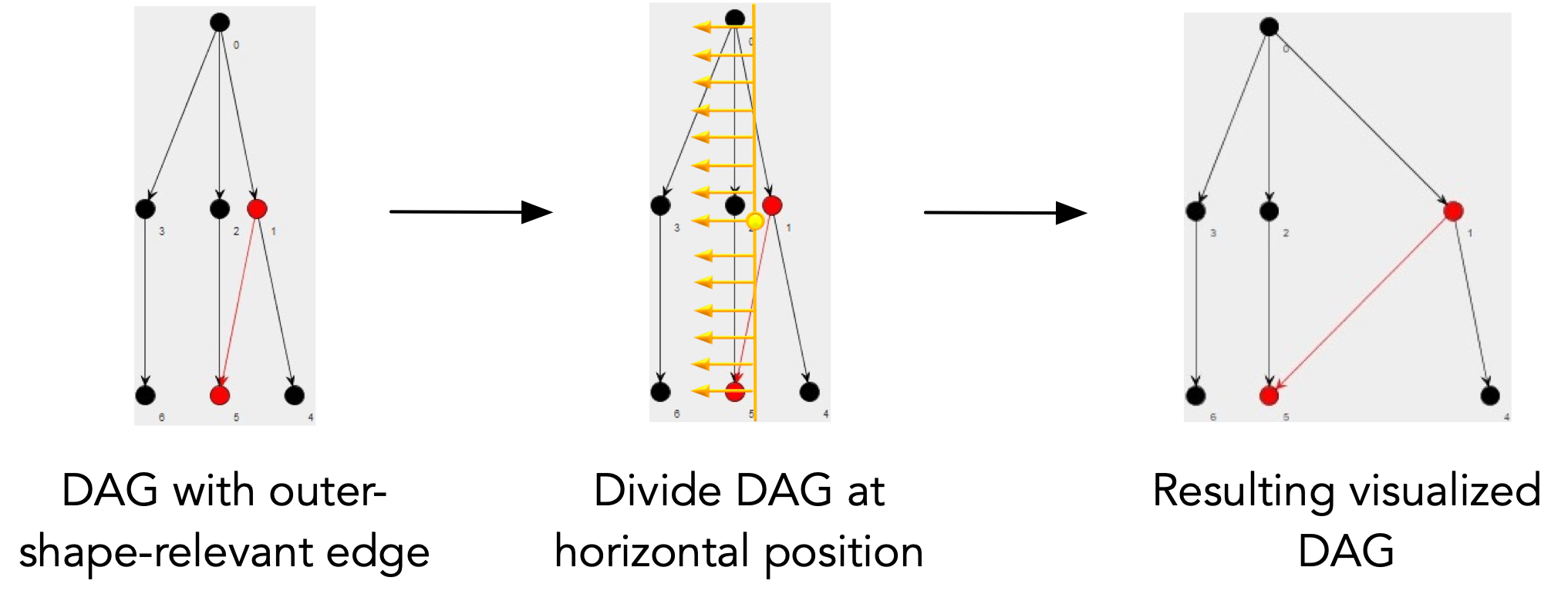}
  \caption[Outer shape change enhancement -- outer-shape-relevant edge change: schematic representation]{Outer shape change enhancement -- outer-shape-relevant edge change: schematic representation.}
  \label{fig:pictures_DACH_ShapeLayout_OuterShapeEnhancement_edge}
\end{figure}

\paragraph{Outer Shape Change Enhancement} 
\label{par:outer_shape_change_enhancement}

Outer shape change enhancements differentiate between nodes and edges and whether they are deleted or added to the DAG. For all adjustments, a precaution is implemented to prevent the average graph aesthetics value from being negatively affected  too much. In case the aesthetics tolerance threshold ($10\%$) is exceeded, the layout is reset to the state of the last iteration.

\textbf{Node Changes.} 
First, the nodes placed on the outside of the DAG are considered (cf. Figure~\ref{fig:pictures_DACH_ShapeLayout_OuterShapeEnhancement_outerNode}). They are suitable for changing the outer shape of the DAG. This approach is the same for deleting and adding nodes. The new or to be deleted node is moved outwards on its horizontal axis: if it has index $0$ of its layer, it is moved to the left. If it has the maximum index of its layer, it is moved to the right. This results in a larger value in the hull difference when adding or deleting a node. The displacement parameter ``adaption'' is configurable. After consulting our participants, we set it to the horizontal node spacing parameter ($80$). If a distinction is not possible because the layer has only one node, the position in relation to the whole DAG is used to decide whether the node is on the left or on the right. An iterative attempt is made to move the node further and further to the left or right until the Hausdorff distance threshold is exceeded. This is the result of a comparison of the base graph's and the alternative graph's concave hull. If the scalar value does not exceed the threshold, the outer shape change is considered imperceptible. In this case, the algorithm is executed again and a before-after hull comparison is performed again. This process can be limited in its number of executions by an upper bound -- here set to $10$. For the Hausdorff distance threshold, we again consulted our participants. The result was that the Hausdorff distance of the hulls must be greater than $\frac{30}{verticalSpacing}=\frac{30}{200}$. Furthermore, this outer shape change enhancement respects the in advanced saved mental map with respect to the nodes' relative position. After this shape change enhancement the nodes' absolute positions are updated. 
%

The next adaptation deals with nodes that are relevant to the outer hull, but do not have the minimum or maximum position index of their layer -- see Figure~\ref{fig:pictures_DACH_ShapeLayout_OuterShapeEnhancement_outerInnerNode} for a schematic representation. From a conversation with a graph layout expert Prof. Dr. A. Kerren, his generalization of magnifying glass interaction with graphs gave rise to the approach of splitting the graph with a vertical axis through the node. As a result, there is an increased hull difference when the node is added or deleted. New edge intersections are also not created, since the relative positions of the nodes to each other were not changed. The graph split induces that all nodes whose horizontal position is to the left of the split center are shifted to the left. The same is true vice versa for the right side. An iterative attempt is made to move the nodes further and further to the left or right until the Hausdorff distance threshold ($\frac{30}{verticalSpacing}=\frac{30}{200}$) is exceeded. This is the result of a comparison of the base graph's and the alternative graph's concave hull. If the scalar value does not exceed the threshold, the outer shape change is considered imperceptible. In this case, the algorithm is executed again and a before-after hull comparison is performed again. This process can be limited in its number of executions by an upper bound -- here set to $10$. This approach is the same for deleting and adding nodes. The displacement parameter ``adaption'' is configurable. After consulting our participants, we set it to the horizontal node spacing parameter ($80$). Furthermore, this outer shape change enhancement respects the in advanced saved mental map with respect to the nodes' relative position. After this shape change enhancement the nodes' absolute positions are updated. 

\textbf{Edge Changes.} There is the possibility to lengthen outer-shape-relevant edges (cf. Figure~\ref{fig:pictures_DACH_ShapeLayout_OuterShapeEnhancement_edge}). In this case, the splitting method is also used. In this process, the split center is positioned at a small distance from the edge origin. It is important that the split center is placed in the direction of the edge destination. An iterative attempt is made to move the nodes left resp. right of the edge destination further and further to the left resp. right until the Hausdorff distance threshold -- $\frac{30}{verticalSpacing}=\frac{30}{200}$ -- is exceeded. This is the result of a comparison of the base graph's and the alternative graph's concave hull. If the scalar value does not exceed the threshold, the outer shape change is considered imperceptible. In this case, the algorithm is executed again and a before-after hull comparison is performed again. This process can be limited in its number of executions by an upper bound -- here set to $10$. This approach is the same for deleting and adding nodes. The displacement parameter ``adaption'' is configurable. After consulting our participants, we set it to the horizontal node spacing parameter ($80$). Furthermore, this outer shape change enhancement respects the in advanced saved mental map with respect to the nodes' relative position. After this shape change enhancement the nodes' absolute positions are updated. 



%



\paragraph{Inner Shape Change Enhancement} 
\label{par:inner_shape_change_enhancement}
\begin{figure}[tb]
   \centering 
   \includegraphics[height=7cm]{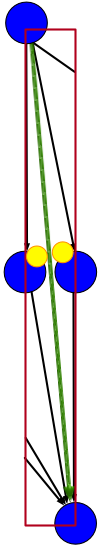}
  \caption[Inner shape change enhancement -- an example of the polygons searched for the inner shape change enhancement]{Inner shape change enhancement -- an example of the polygons searched for the inner shape change enhancement: the polygon with the largest inner circle which is adjacent to the DAG change and the respective polygon on the opposite site of the one with the largest inner circle. In this example there are only to polygons adjacent to the change. However, there may be DAG changes where there are significantly more polygons adjacent to the DAG's change. (Figure based on Figure from \cite{MarcMA})}
  \label{fig:pictures_DACH_ShapeLayout_WhiteSpaceCaclulation_RectangleEnclosesChange2}
\end{figure}

\begin{figure}[tb]
  \centering
    \includegraphics[width=.9\textwidth]{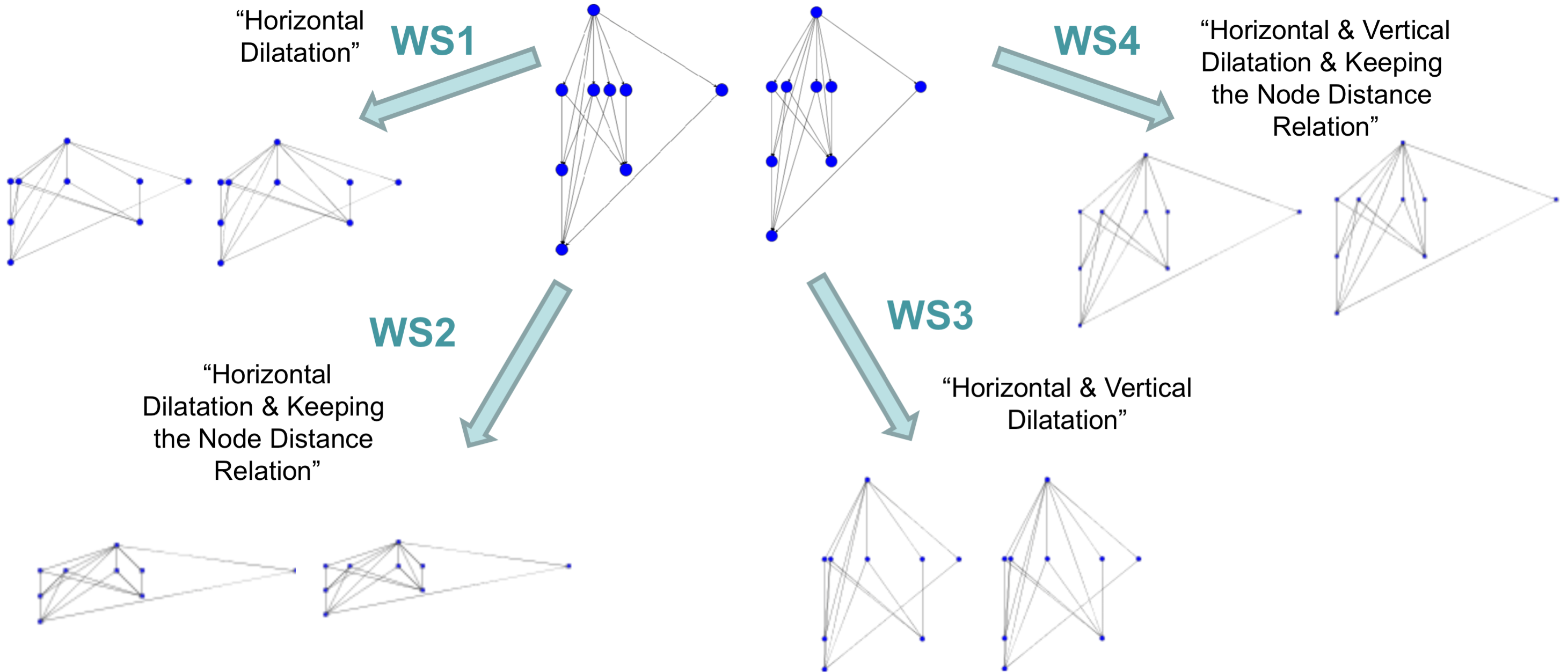}
  \caption[Inner shape change enhancement -- approach variants]{Inner shape change enhancement -- approach variants: Already here we can spot strength and weaknesses of the different approaches. For example, while the approaches WS1 and WS3 nicely enlarge the white space around the added edge, the approaches WS2 and WS4 rather lead to white space on the right next to the added edge.}
  \label{fig:pictures_DACH_ShapeLayout_WhiteSpaceApproaches}
\end{figure}

For the inner shape change enhancements we employ our inner shape algorithm (cf. Section \texttt{Inner Shape -- White Space}). Here, however, we do not search all inner circles of the polygons in the bounding rectangle. We search for the polygon with the largest inner circle which is adjacent to the DAG change (cf. Figure~\ref{fig:pictures_DACH_ShapeLayout_WhiteSpaceCaclulation_RectangleEnclosesChange2}). Further, we search for the polygon on the opposite site of the one with the largest inner circle which is adjacent to the DAG change (cf. Figure~\ref{fig:pictures_DACH_ShapeLayout_WhiteSpaceCaclulation_RectangleEnclosesChange2}). We search for the polygon with the largest inner circle as enlarging the largest inner circle and the on the opposite side has a strong enlarging effect on the DAGs' white space without distorting the visualized DAG entirely.

\textbf{Node and Edge Changes.} It depends on whether it is a node or edge change how the bounding rectangle is defined. If the bounding rectangle encloses an edge change, as in Figure~\ref{fig:pictures_DACH_ShapeLayout_WhiteSpaceCaclulation_RectangleEnclosesChange2}, then the width of the bounding rectangle for edges is defined by the coordinates of the nodes that this edge connects. For a node change, the width of the bounding box is defined by the nodes' positions adjacent to the changing node and remaining the same.

\textbf{Approaches.} We propose four different inner shape change enhancement approaches (WS1-WS4). Figure~\ref{fig:pictures_DACH_ShapeLayout_WhiteSpaceApproaches} shows a visualization of one and the same DAG pair visualized based on the four different white space enlarging approaches. Already here, in Figure~\ref{fig:pictures_DACH_ShapeLayout_WhiteSpaceApproaches}, we can spot strength and weaknesses of the different approaches. For example, while the approaches WS1 and WS3 nicely enlarge the white space around the added edge, the approaches WS2 and WS4 rather lead to white space on the right next to the added edge.\\
\emph{\textbf{WS1 -- Horizontal Shift.}} WS1 shifts the nodes only on the horizontal layer, starting from the center of the inner circle. Nodes with a lower x-coordinate than the center of the inner circle are shifted to the left and those with a higher x-coordinate are shifted to the right. The displacement parameter ``adaption''  is configurable. After consulting our participants, we set this to twice the horizontal node spacing parameter. The displacement is multiplied by the factor f. It indicates how much of the actual displacement potential -- displacement parameter ``adaptation'' -- the circle receives. The smaller circle is scaled much higher because the size adjustment of both is aimed at -- i.e., both circle shall be of equal size. 
\emph{\textbf{WS2 -- Horizontal Shift \& Keeping the Node Distance Ratio.}} The second variant, like the first, allows only horizontal shifts. However, the nodes are moved in relation to their previous distance from the center of the inner circle, and not absolutely, as in WS1. 
\emph{\textbf{WS3 -- Horizontal \& Vertical Shift.}} WS3 shifts the nodes on the horizontal and vertical layer, starting from the center of the inner circle. Nodes with a lower x-coordinate than the center of the inner circle are shifted to the left and those with a higher x-coordinate are shifted to the right. Nodes with a lower y-coordinate than the center coordinate are shifted to the top and those with a higher one are shifted to the bottom. The displacement parameter ``adaption''  is also configurable here. As we approximate the white space with the largest inner circle of the polygons, we have to shift the nodes on the x- and the y-axis using the same displacement parameter configuration -- otherwise we would create an ellipse. Here, we also use the parameter setting of WS1. Again, the displacement is multiplied by the factor f. 
\emph{\textbf{WS4 -- Horizontal \& Vertical Shift \& Keeping the Node Distance Ratio.}} The fourth variant also allows horizontal and vertical shifts. However, unlike the third variant, the nodes are moved in relation to their previous distance from the center of the inner circle, and not absolutely, as in WS1 and WS3. 

In all approaches, the circles are enlarged until the white space areas of the two circles exceed the white space threshold. This threshold is relative to the graph area. Consultation with our participants indicated that the white space threshold should be larger than $5\%$ of the total graph area of the visualized DAG. We calculate the graph area based on the convex hull. Similarly, for all approaches, a precaution is implemented to prevent the average graph aesthetics value from being negatively affected too much. In case the aesthetics tolerance threshold ($10\%$) is exceeded, the layout is reset to the state of the last iteration. Furthermore, this outer shape change enhancement respects the in advanced saved mental map with respect to the nodes' relative position. The iterative attempt to enlarge the white space can be limited in its number of executions by an upper bound -- here set to $10$.
\FloatBarrier

\subsubsection{Post-Processing} 
\label{ssub:post_processing}
The resulting layout applies to the supergraph. To use the layout for the base graph, the supergraph is rebuilt to resemble the base graph. The rebuilt supergraph then replaces the base graph. The supergraph is then rebuilt to match the alternative graph. This then replaces the alternative graph. This results in the layout of the alternative graph.


\subsection{Evaluation} 
\label{sub:evaluation}
In our evaluation, we compare our shape change enhancing layout with our base implementation which we explained in Section~\ref{sub:our_shape_change_enhancing_hierarchical_layout_algorithm}. For our shape change enhancing layout, we evaluate our outer shape change enhancements and the different approaches we proposed for the inner shape change enhancement. For the inner ones, however, we leave out approach WS2 as it performs considerably worse than the others. Figure~\ref{fig:pictures_DACH_ShapeLayout_WhiteSpaceApproaches} substantiates that.

\subsubsection{Dataset and Graph Changes} 
\label{ssub:dataset_and_graph_chnages}
We randomly generated $100$ base graphs with $21$ nodes and $61$ edges. This corresponds to realistic size and properties of for instance contagion graphs of hospital disease spread or gene mutations~\cite{v.20191328,lenz2014visual}. We change the base graphs by adding/deleting one node/edge, by adding three nodes, and by adding one node and two edges. We perform those changes at all possible positions in the base graph.

We first analyze each change individually. The change types are: added node, removed node, added edge, removed edge. The advantage of the single changes is that hull-relevant changes and white space changes can be considered and evaluated in isolation without interacting with each other. The evaluation of the single change is the controlled setting to see if the chosen approach works per se. 
 We investigate the performance of our layout for multiple simultaneous changes using the following two scenarios: added $3$ nodes, added $1$ node and $2$ edges.

 These are realistic in different contexts -- explained using the hospitalization context: 
 \begin{itemize}
 	\item Added $3$ nodes:\\
	It may well be that three patients are hospitalized in one day.
 	\item Added $1$ node and $2$ edges:\\
	Another admitted patient is placed in a three-bed room. There are already two other patients there. It may also be that one patient is hospitalized in a single room and two pairs of patients have met at a checkup.
 \end{itemize}


\subsubsection{Evaluation Metrics} 
\label{ssub:evaluation_metrics}
We employ specific metrics for the outer-shape-relevant changes and for the inner-shape-relevant changes. The graph aesthetic, we employ for both the outer and the inner shape change enhancements since the graph aesthetics are the fundamental rules of graph drawing which aught not to be negatively impacted by both of our shape change enhancements. We also calculate the absolute difference, the ratio, and the (weighted) average for each shape change enhancement evaluation metric.

\paragraph{Outer Shape Change Enhancement Evaluation Metrics.} 
\label{par:outer_shape_change_enhancement_evaluation_metrics}
\begin{figure}[tb]
  \centering
    \includegraphics[width=.5\textwidth]{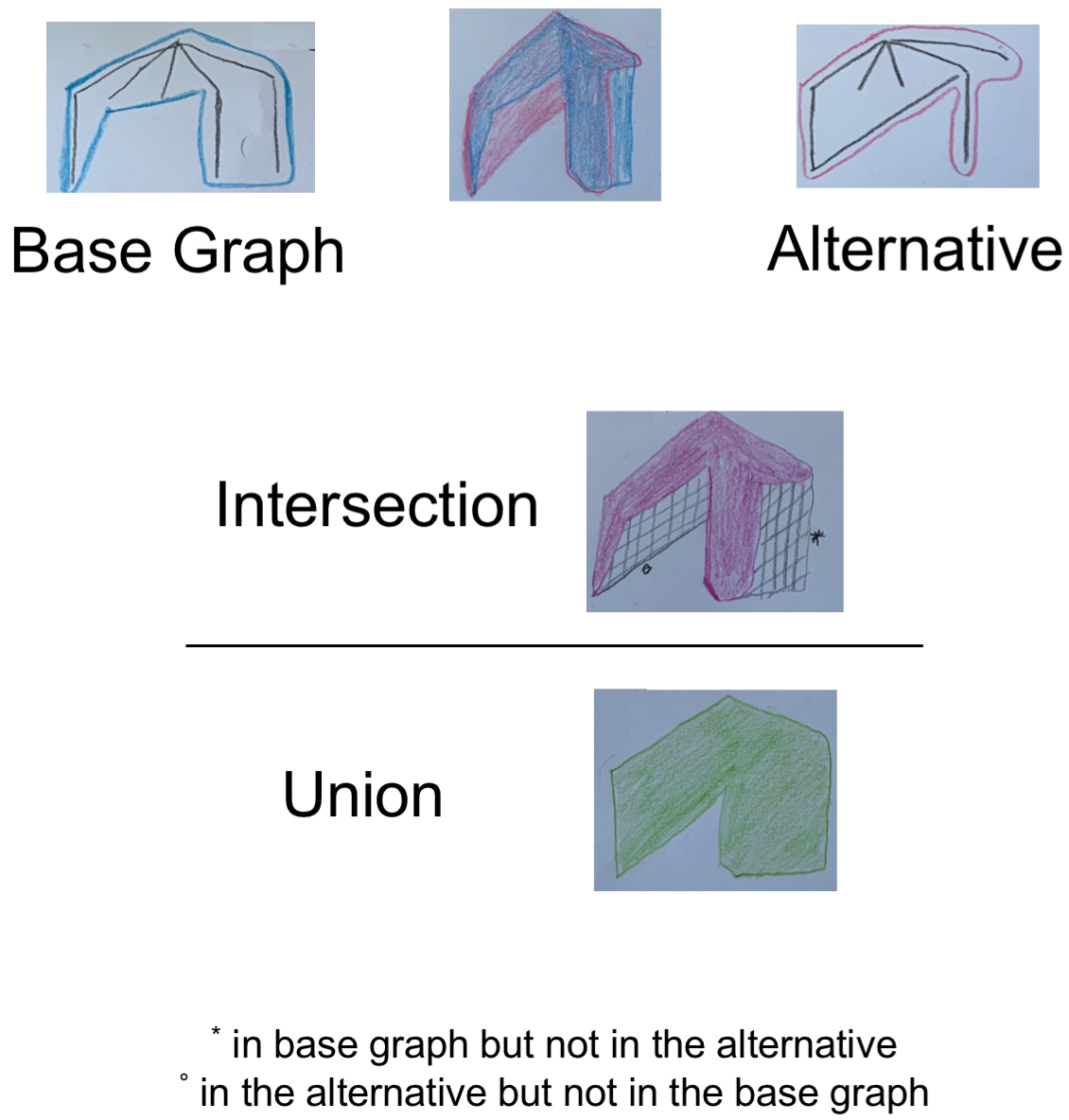}
  \caption[Shape change enhancement evaluation metric -- intersection over union: schematic representation]{Shape change enhancement evaluation metric -- intersection over union: schematic representation.}
  \label{fig:pictures_DACH_ShapeLayout_IoU}
\end{figure}

\textbf{Normalized Hausdorff Distance Average.} As Huttenlocher et al. \cite{232073} explain, the Hausdorff distance measures the extent to which each point of [one] set lies near some point of [another] set and vice versa. Consequently, it is a good and easy and quickly to calculate approximation of how much the outer hull of two DAGs changed. For the formula and calculation details, please refer to \cite{6849454}. To improve comparability, we also take the hull size into account and normalize the Hausdorff distance results by multiplying it with $\frac{1}{concaveHullSize}$. We calculate the average by summing up all the normalized Hausdorff distance results for all alternatives and dividing the sum by the number of alternatives.

Since the Hausdorff distance works with the maximum, it would only consider the maximum hull change when there are more then one DAG change which are outer-shape-relevant. In our point of view, this would be an approximation which is too rough. So, we introduce the intersection over union metric for more than one outer shape change enhancement.

The \textbf{Intersection over Union} is the average ratio of the intersection area of the concave hull of the base graph and the concave hull of the alternative graph in relation to the union area of these. An exemplary illustration is shown in Figure~\ref{fig:pictures_DACH_ShapeLayout_IoU} below, where the hull of a base graph is shown on the left and the hull of an alternate graph is shown on the right. In the center, both hulls are superimposed. There you can see the intersection area. It is the area where the blue hull of the base graph and the pink hull of the alternative graph overlap (cf. also Figure~\ref{fig:pictures_DACH_ShapeLayout_IoU} -- \texttt{Intersection}). The union area is the area which is covered by both hulls (cf. Figure~\ref{fig:pictures_DACH_ShapeLayout_IoU} -- \texttt{Union}). The intersection over union result expresses the area change in relation to the area covered in total.
\FloatBarrier

\paragraph{Inner Shape Change Enhancement Evaluation Metrics.} 
\label{par:inner_shape_change_enhancement_evaluation_metrics}

\begin{figure}[htb]
  \centering
    \includegraphics[width=.4\textwidth]{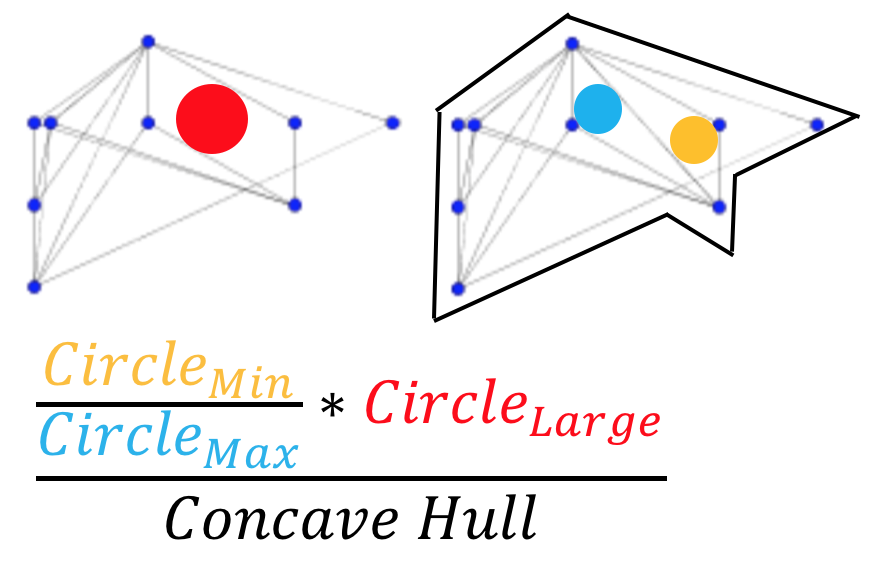}
  \caption[Shape change enhancement evaluation metric -- relative white space average: schematic representation]{Schematic representation of the relative white space average evaluation metric. For each inner shape change, the resulting white space is approximated by its inner circle (\redLargeCircle{$\mathtt{Circle_{Large}}$}). This is multiplied by the ratio of the two white space circles adjacent to the inner change (\yellowMinCircle{$\mathtt{Circle_{Min}}$}, \blueMaxCircle{$\mathtt{Circle_{Max}}$}). Here, the assumption is that a change is seen particularly well if the two circles adjacent to the inner change are as equal as possible. The change is then as central as possible in the white space. For comparability of the results between the graphs, the result is divided by the concave hull (\texttt{Concave Hull}).}
  \label{fig:pictures_DACH_ShapeLayout_RelativeWhiteSpaceAverage}
\end{figure}

\textbf{Relative White Space Average.} 
Figure~\ref{fig:pictures_DACH_ShapeLayout_RelativeWhiteSpaceAverage} shows a schematic of the formula. For each inner shape change, the resulting white space is approximated by its inner circle (cf. Figure~\ref{fig:pictures_DACH_ShapeLayout_RelativeWhiteSpaceAverage} -- \redLargeCircle{$\mathtt{Circle_{Large}}$}). This is multiplied by the ratio of the two white space circles adjacent to the inner change -- \yellowMinCircle{$\mathtt{Circle_{Min}}$}, \blueMaxCircle{$\mathtt{Circle_{Max}}$} -- because the assumption is that a change is seen particularly well if the two circles adjacent to the inner change are as equal as possible. The change is then as central as possible in the white space. For comparability of the results between the graphs, the result is divided by the concave hull (Figure~\ref{fig:pictures_DACH_ShapeLayout_RelativeWhiteSpaceAverage} -- \texttt{Concave Hull}). We calculate the average by summing up all the relative white spaces for all alternatives and dividing the sum by the number of alternatives.


\paragraph{Evaluation Metrics for Both Shape Change Enhancements} 
\label{par:evaluation_metrics_for_both_shape_change_enhancements}

\textbf{Aesthetic Criteria Average.} For all alternative graphs, all previously defined aesthetics criteria were calculated and the average of these was formed under equal weighting. A value of $1.0$ means that all alternatives in each criterion have an aesthetics value of $1.0$. The aesthetics we use are those from H. Purchase \cite{purchase2002metrics} and Dunne et al. \cite{10.1147/JRD.2015.2411412} since the current body of work on drawing quality evaluation reflects that those are amongst the most frequently used criteria -- also specifically for our graph and visualization type. The aesthetics criteria of H. Purchase and Dunne et al. i.a. encompass the number of edge crossings, angular resolution, edge bend, and symmetry. For the encompassing list and details on the aesthetics' calculation please refer to the work of H. Purchase \cite{purchase2002metrics} and Dunne et al. \cite{10.1147/JRD.2015.2411412}. Currently, we use equal weights for all aesthetic criteria, since it is not apparent from related work which aesthetic criterion should be weighted more heavily than another. We introduced a weighting factor for each aesthetic criterion to allow a weighting when the related work does provide this information in the future or the allow the user of our layout to configure it according to her needs. The criterion is needed to calculate how high the costs of our shape change enhancements are in relation to the drawing quality which is measured by the aesthetic criteria. We calculate the average by summing up all the aesthetic criteria results for all alternatives and dividing the sum by the number of alternatives.

\textbf{Absolute Difference of Our Shape Change Enhancing Layout and of the Basis Implementation.} For a difference column, each is the absolute average difference between the respective computed metric for our shape change enhancing layout and for the base implementation. For example, the difference column of the average Hausdorff distance indicates the absolute average difference between the average Hausdorff distance of our layout and that of the base implementation. The difference tells the absolute average by how much better or worse our layout is.

\textbf{Better, Equal, or Worse in Percent.} This metric describes the percentage relative to the number of all alternatives of how often our layout was better than the base implementation, how often both were equally good, and how often our layout was worse than the base implementation. The calculation of these values is based on the absolute difference metric.\\
\emph{\textbf{Basis Implementation Better.}} In cases when the base implementation outperforms our layout, the absolute difference is $<0$.\\
\emph{\textbf{Basis Implementation = Shape Change Enhancing Layout.}} If both layouts are equally good, then the absolute difference is $0$.\\
\emph{\textbf{Shape Change Enhancing Layout Better.}} In case our layout outperforms the base implementation, the absolute difference is $>0$.

\textbf{Metric Ratio.} The ratio column is formed by dividing the metrics results for our layout by those for the base implementation. This results in the respective average ratio of the metric results of our layout to the results of the base implementation. The ratio of our layout to the base implementation indicates how much better or worse our layout is.

\textbf{Average.} When calculating the the average of a metric, each metric result is equally weighted.\\
\emph{\textbf{Weighted Average.}} When calculating the weighted average of a metric, each metric result is weighted by the respective dataset proportion. The dataset proportion here is the proportion of DAGs that the respective change type produces. The change types are: added $1$ node, added $1$ edge, deleted $1$ node, deleted $1$ edge, added $3$ nodes, added $1$ node and $2$ edges.
Both type of averages indicate how the performance of the respective layout over all alternatives and change types is. So, we can gain an impression of the overall layout performance.

\subsubsection{Evaluation Results} 
\label{ssub:evaluation_results}
The presentation of our evaluation results is structured as follows: We first present the results of the single changes -- add/delete $1$ node/edge. There, we first elaborate on the results of the outer shape change enhancement evaluation and second on the inner shape change enhancement evaluation results. After that we present our results for multiple simultaneous changes. For those, we also first present the outer shape change enhancement results and then the inner shape change enhancement results.

\paragraph{Single Changes -- Outer Shape Change Enhancement Evaluation Results, Discussion, and Conclusion.} 
\label{par:outer_shape_change_enhancement_results}

\begin{table}[H]
	\centering
\begin{tabular}{|l|l|l|l|l|l|}
\hline
                                       & \parbox[t][][t]{0.15\textwidth}{\raggedright \textbf{Hausdorff Distance Average (Base Implementation)}} & \parbox[t][][t]{0.15\textwidth}{\raggedright \textbf{Hausdorff Distance Average (Shape Change Enhancing Layout)}} & \parbox[t][][t]{0.12\textwidth}{\textbf{Absolute Difference}} & \parbox[t][][t]{0.12\textwidth}{\textbf{Metric Ratio}} & \parbox[t][][t]{0.15\textwidth}{\textbf{Dataset \\Proportion}} \\ \hline
\parbox[t][][t]{0.15\textwidth}{\textbf{Add 1 Edge}} &  $0.2094$                                              &  $0.5842$                                   &  $0.3748$                              &  $2.7904$             &  $0.4085$                   \\ \hline
\parbox[t][][t]{0.15\textwidth}{\textbf{Add 1 Node}} &  $0.7097$                                              &  $1.1822$                                   &  $0.4725$                              &  $1.6658$             &  $0.4478$                   \\ \hline
\parbox[t][][t]{0.15\textwidth}{\raggedright \textbf{Remove 1 Edge}} &  $0.1648$                              &  $0.3920$                                   &  $0.2272$                              &  $2.3779$             &  $0.1076$                   \\ \hline
\parbox[t][][t]{0.15\textwidth}{\raggedright \textbf{Remove 1 Node\textcolor{white}{g}}} &  $0.6381$          &  $0.9992$                                   &  $0.3611$                              &  $1.5659$             &  $0.0361$                   \\ \hline \hline
\parbox[t][][t]{0.15\textwidth}{\textbf{Equally Weighted Average}} &  $0.4305$                                &  $0.7894$                                   &  $0.3589$                              &  $2.1000$             &                                        \\ \hline
\parbox[t][][t]{0.15\textwidth}{\raggedright \textbf{Average Weighted by Dataset Proportion}} &  $0.4441$     &  $0.8463$                                   &  $0.4022$                              &  $2.1982$             &                                        \\ \hline
\end{tabular}
\caption[Single changes -- outer shape change enhancements: normalized Hausdorff distance average results]{Single changes -- outer shape change enhancements: normalized Hausdorff distance average results.}
\label{tab:tabelle6}
\end{table}

\begin{table}[H]
	\centering
\begin{tabular}{|l|l|l|l|l|}
\hline
                                       & \parbox[t][][t]{0.15\textwidth}{\raggedright \textbf{Base Implementation Better}} & \parbox[t][][t]{0.15\textwidth}{\raggedright \textbf{Base Implementation = Shape Change Enhancing Layout}} & \parbox[t][][t]{0.12\textwidth}{\textbf{Shape Change Enhancing Layout Better}} & \parbox[t][][t]{0.15\textwidth}{\textbf{Dataset \\Proportion}} \\ \hline
\parbox[t][][t]{0.15\textwidth}{\textbf{Add 1 Edge}} &  $0.0185$                                              &  $0.2904$                                                         &  $0.6910$                                                           &  $0.4085$ \\ \hline
\parbox[t][][t]{0.15\textwidth}{\textbf{Add 1 Node}} &  $0.0018$                                              &  $0.2134$                                                         &  $0.7848$                                                           &  $0.4478$ \\ \hline
\parbox[t][][t]{0.15\textwidth}{\raggedright \textbf{Remove 1 Edge}} &  $0.0555$                              &  $0.2774$                                                         &  $0.6671$                                                           &  $0.1076$ \\ \hline
\parbox[t][][t]{0.15\textwidth}{\raggedright \textbf{Remove 1 Node\textcolor{white}{g}}} &  $0.0074$          &  $0.3603$                                                         &  $0.6324$                                                           &  $0.0361$ \\ \hline \hline
\parbox[t][][t]{0.15\textwidth}{\textbf{Equally Weighted Average}} &  $0.0208$                                &  $0.2854$                                                         &  $0.6938$                                                           &           \\ \hline 
\parbox[t][][t]{0.15\textwidth}{\raggedright \textbf{Average Weighted by Dataset Proportion}} &  $0.0146$    &  $0.2571$                                                         &  $0.7283$                                                           &           \\ \hline
\end{tabular}
\caption[Single changes -- outer shape change enhancements: better, equal, or worse in percent]{Single changes -- outer shape change enhancements: better, equal, or worse in percent results with respect to the Hausdorff distance average metric.}
\label{tab:tabelle8}
\end{table}

\begin{table}[H]
	\centering
\begin{tabular}{|l|l|l|l|l|l|l|l|}
\hline
                                       & \parbox[t][][t]{0.1\textwidth}{\raggedright \textbf{Aes-\\thetic Criteria Average (Base Implementation)}} & \parbox[t][][t]{0.1\textwidth}{\raggedright \textbf{Aes-\\thetic Criteria Average (Shape Change Enhancing Layout)}} & \parbox[t][][t]{0.1\textwidth}{\textbf{Aes-\\thetic Criteria Average Absolute Difference}} & \parbox[t][][t]{0.1\textwidth}{\textbf{Aes-\\thetic Criteria Average Metric Ratio}} & \parbox[t][][t]{0.10\textwidth}{\raggedright \textbf{Edge Crossings Average (Base Implementation)}} & \parbox[t][][t]{0.10\textwidth}{\raggedright \textbf{Edge Crossings Average (Shape Change Enhancing Layout)}} & \parbox[t][][t]{0.15\textwidth}{\textbf{Dataset \\Proportion}} \\ \hline
\parbox[t][][t]{0.10\textwidth}{\raggedright \textbf{Add 1 Edge}} &  $0.6734$                                 & $0.6319$                     & $0.0416$                                    & $0.9382$              & $171.9860$              & $172.6732$             & $0.4085$                   \\ \hline
\parbox[t][][t]{0.10\textwidth}{\raggedright \textbf{Add 1 Node\textcolor{white}{g}}} &  $0.6699$             & $0.6587$                     & $0.0111$                                    & $0.9834$              & $165.4487$              & $165.7475$             & $0.4478$                   \\ \hline
\parbox[t][][t]{0.10\textwidth}{\raggedright \textbf{Remove 1 Edge}} & $0.6734$                               & $0.6487$                     & $0.0248$                                    & $0.9632$              & $160.2947$              & $161.0764$             & $0.1076$                    \\ \hline
\parbox[t][][t]{0.10\textwidth}{\raggedright \textbf{Remove 1 Node\textcolor{white}{g}}} & $0.6727$           & $0.6677$                     & $0.0050$                                    & $0.9926$              & $139.0258$              & $139.3432$             & $0.0361$                   \\ \hline \hline
\parbox[t][][t]{0.15\textwidth}{\textbf{Equally Weighted Average}} & $0.6724$                                 & $0.6517$                     & $0.0206$                                    & $0.9694$              & $159.1888$              & $159.7101$             &                            \\ \hline 
\parbox[t][][t]{0.15\textwidth}{\raggedright \textbf{Average Weighted by Data-\\set Proportion}} & $0.6718$   & $0.6470$                     & $0.0248$                                    & $0.9631$              & $166.6106$              & $167.1207$             &                            \\ \hline
\end{tabular}
\caption[Single changes -- outer shape change enhancements: aesthetic criteria average and edge crossings result]{Single changes -- outer shape change enhancements: aesthetic criteria average and edge crossings result.}
\label{tab:tabelle11}
\end{table}

%
%
%

\textbf{Normalized Hausdorff Distance Average.} 
For the average Hausdorff distance, it is noticeable that our shape change enhancing layout is about twice as good as the base implementation for all values (cf. Table~\ref{tab:tabelle6}). This is true for the comparison of the absolute values of the average normalized Hausdorff distance for the different change types, as well as for the absolute difference values and the ratio values. The equally weighted average and the average weighted by dataset proportion confirm the impression that can be gained from the other values. It is also noticeable that the optimization is positive especially for the edge operations (cf. Table~\ref{tab:tabelle6} -- \texttt{ratio}: $2.79$ and $2.38$). There, the relative ratios are particularly high compared to the base implementation. This is probably because nodes inherently produce more shape difference.\\
Relative to the average weighted by the dataset proportion, the base implementation performs better than our layout in only $1.46\%$ of all alternatives and equally well in $25.71\%$ of all alternatives (cf. Table~\ref{tab:tabelle8}). The equal or better performance of the base implementation is due to the fact that our shape change enhancements are reset if they exceed the tolerance parameters, such as the one for the aesthetics criteria. In $72.83\%$ of all alternatives our layout outperforms the base implementation (cf. Table~\ref{tab:tabelle8}).

\textbf{Aesthetic Criteria Average.} The evaluation of the aesthetics criteria average and specifically the number of edge crossings -- see Table~\ref{tab:tabelle11} -- reveals: According to the weighted average by dataset proportion, our shape change enhancements produce a loss of a mere $3.70\%$ in the aesthetics criteria average. This is a very good result for the rather intensive interventions performed by our shape change enhancements and the increase in the number of outer-shape-relevant graph elements. We set the parameter to $10\%$ allowed loss for the aesthetic criteria average. Looking at the individual losses, we clearly see that the $10\%$ average is far from being reached -- add $1$ edge: $6.18\%$; add $1$ node: $1.66\%$; delete $1$ edge: $3.68\%$, add $1$ node: $0.74\%$. Edges tend to have a higher impact on the aesthetics criteria, as edges have an impact on edge crossings, edge angles, edge length, and more. Looking specifically at edge crossings, we can see that our layout only produces about $0.5$ more edge crossings on average than the base implementation for the outer shape change enhancements. So this is a negligible increase in edge crossings.

\textbf{Conclusion.} Our layout convinces with its outer shape change enhancements. It outperforms the base implementation in $72.83\%$ of all alternatives and the cost in terms of aesthetics criteria is only $3.70\%$. This shows that the optimization functions achieve their goal.

\FloatBarrier


\paragraph{Single Changes -- Inner Shape Change Enhancement Evaluation Results, Discussion, and Conclusion.} 
\label{par:single_changes_inner_shape_change_enhancement_evaluation_results}

\begin{table}[H]
	\centering
\begin{tabular}{|l|l|l|l|l|l|}
\hline
                                       & \parbox[t][][t]{0.15\textwidth}{\raggedright \textbf{Relative White Space Average (Base Implementation)}} & \parbox[t][][t]{0.2\textwidth}{\raggedright \textbf{Relative White Space Average (Shape Change Enhancing Layout)}} & \parbox[t][][t]{0.12\textwidth}{\textbf{Absolute Difference}} & \parbox[t][][t]{0.12\textwidth}{\textbf{Metric Ratio}} & \parbox[t][][t]{0.15\textwidth}{\textbf{Dataset \\Proportion}} \\ \hline
\parbox[t][][t]{0.2\textwidth}{\raggedright \textbf{Add 1 Edge (WS1)}} &  $0.0016$                                              & $0.0054$                                    & \textcolor{white}{$-$}$0.0038$                               & $3.3082$              & $0.6178$                    \\ \hline
\parbox[t][][t]{0.2\textwidth}{\raggedright \textbf{Add 1 Edge (WS3)}} &  $0.0017$                                              & $0.0065$                                    & \textcolor{white}{$-$}$0.0048$                               & $3.9108$              & $0.6238$                    \\ \hline
\parbox[t][][t]{0.2\textwidth}{\raggedright \textbf{Add 1 Edge (WS4)}} &  $0.0017$                                              & $0.0019$                                    & \textcolor{white}{$-$}$0.0003$                               & $1.1721$              & $0.6166$                    \\ \hline
\parbox[t][][t]{0.2\textwidth}{\raggedright \textbf{Add 1 Node (WS1)}} &  $0.0015$                                              & $0.0034$                                    & \textcolor{white}{$-$}$0.0019$                               & $2.2296$              & $0.0278$                    \\ \hline
\parbox[t][][t]{0.2\textwidth}{\raggedright \textbf{Add 1 Node (WS3)}} &  $0.0021$                                              & $0.0024$                                    & \textcolor{white}{$-$}$0.0003$                               & $1.1566$              & $0.0191$                    \\ \hline
\parbox[t][][t]{0.2\textwidth}{\raggedright \textbf{Add 1 Node (WS4)}} &  $0.0017$                                              & $0.0016$                                    & $-0.0001$                                                    & $0.9234$              & $0.0209$                    \\ \hline
\parbox[t][][t]{0.2\textwidth}{\raggedright \textbf{Remove 1 Edge (WS1)}} & $0.0016$                                            & $0.0032$                                    & \textcolor{white}{$-$}$0.0015$                               & $1.9449$              & $0.3542$                    \\ \hline
\parbox[t][][t]{0.2\textwidth}{\raggedright \textbf{Remove 1 Edge (WS3)}} & $0.0017$                                            & $0.0044$                                    & \textcolor{white}{$-$}$0.0027$                               & $2.6279$              & $0.3568$                    \\ \hline
\parbox[t][][t]{0.2\textwidth}{\raggedright \textbf{Remove 1 Edge (WS4)}} & $0.0017$                                            & $0.0020$                                    & \textcolor{white}{$-$}$0.0003$                               & $1.1833$              & $0.3625$                    \\ \hline
\parbox[t][][t]{0.2\textwidth}{\raggedright \textbf{Remove 1 Node (WS1)\textcolor{white}{g}}} & $0.0001$                        & $0.0001$                                    & $-0.00001$                                                   & $0.9107$              & $0.0001$                    \\ \hline
\parbox[t][][t]{0.2\textwidth}{\raggedright \textbf{Remove 1 Node (WS3)\textcolor{white}{g}}} & $0.0001$                        & $0.0007$                                    & \textcolor{white}{$-$}$0.0006$                               & $7.2210$              & $0.0003$                    \\ \hline
\parbox[t][][t]{0.2\textwidth}{\raggedright \textbf{Remove 1 Node (WS4)\textcolor{white}{g}}} & $0.0000$                        & $0.0000$                                    & \textcolor{white}{$-$}$0.0000$                               & $1.0000$              & $0.000001$                    \\ \hline \hline
\parbox[t][][t]{0.2\textwidth}{\raggedright \textbf{Eq. Weighted Avg. (WS1)}} & $0.0012$                                 & $0.0030$                                    & \textcolor{white}{$-$}$0.0018$                               & $1.0984$              &                                        \\ \hline
\parbox[t][][t]{0.2\textwidth}{\raggedright \textbf{Eq. Weighted Avg. (WS3)}} & $0.0014$                                 & $0.0035$                                    & \textcolor{white}{$-$}$0.0021$                               & $3.7290$              &                                        \\ \hline
\parbox[t][][t]{0.2\textwidth}{\raggedright \textbf{Eq. Weighted Avg. (WS4)}} & $0.0013$                                 & $0.0014$                                    & \textcolor{white}{$-$}$0.0021$                               & $1.0697$              &                                        \\ \hline
\parbox[t][][t]{0.2\textwidth}{\raggedright \textbf{Avg. Weighted by Dataset Prop. (WS1)}} & $0.0016$                   & $0.0046$                                    & \textcolor{white}{$-$}$0.0029$                               & $2.7949$              &                                        \\ \hline
\parbox[t][][t]{0.2\textwidth}{\raggedright \textbf{Avg. Weighted by Dataset Prop. (WS3)}} & $0.0017$                   & $0.0057$                                    & \textcolor{white}{$-$}$0.0040$                               & $3.4015$              &                                        \\ \hline
\parbox[t][][t]{0.2\textwidth}{\raggedright \textbf{Avg. Weighted by Dataset Prop. (WS4)}} & $0.0017$                   & $0.0019$                                    & \textcolor{white}{$-$}$0.0003$                               & $1.1710$              &                                        \\ \hline
\end{tabular}
\caption[Single changes -- inner shape change enhancements: relative white space average results]{Single changes -- inner shape change enhancements: relative white space average results.}
\label{tab:tabelle14}
\end{table}

\begin{table}[H]
	\centering
\begin{tabular}{|l|l|l|l|l|}
\hline
                                       & \parbox[t][][t]{0.15\textwidth}{\raggedright \textbf{Base Implementation Better}} & \parbox[t][][t]{0.15\textwidth}{\raggedright \textbf{Base Implementation = Shape Change Enhancing Layout}} & \parbox[t][][t]{0.12\textwidth}{\textbf{Shape Change Enhancing Layout Better}} & \parbox[t][][t]{0.15\textwidth}{\raggedright \textbf{Dataset \\Proportion}} \\ \hline
\parbox[t][][t]{0.2\textwidth}{\raggedright \textbf{Add 1 Edge (WS1)}} &  $0.3291$                                              &  $0.0315$                                                         &  $0.6394$                                                           &  $0.6178$ \\ \hline
\parbox[t][][t]{0.2\textwidth}{\raggedright \textbf{Add 1 Edge (WS3)}} &  $0.2991$                                              &  $0.0692$                                                         &  $0.6318$                                                           &  $0.6238$ \\ \hline
\parbox[t][][t]{0.2\textwidth}{\raggedright \textbf{Add 1 Edge (WS4)}} &  $0.3674$                                              &  $0.0587$                                                         &  $0.5740$                                                           &  $0.6166$ \\ \hline
\parbox[t][][t]{0.2\textwidth}{\raggedright \textbf{Add 1 Node (WS1)}} &  $0.4321$                                              &  $0.0309$                                                         &  $0.5340$                                                           &  $0.0278$ \\ \hline
\parbox[t][][t]{0.2\textwidth}{\raggedright \textbf{Add 1 Node (WS3)}} &  $0.5683$                                              &  $0.0573$                                                         &  $0.3744$                                                           &  $0.0191$ \\ \hline
\parbox[t][][t]{0.2\textwidth}{\raggedright \textbf{Add 1 Node (WS4)}} &  $0.5061$                                              &  $0.1020$                                                         &  $0.3918$                                                           &  $0.0209$ \\ \hline
\parbox[t][][t]{0.2\textwidth}{\raggedright \textbf{Remove 1 Edge (WS1)}} &  $0.4418$                                           &  $0.0332$                                                         &  $0.5251$                                                           &  $0.3542$ \\ \hline
\parbox[t][][t]{0.2\textwidth}{\raggedright \textbf{Remove 1 Edge (WS3)}} &  $0.3027$                                           &  $0.0779$                                                         &  $0.6194$                                                           &  $0.3568$ \\ \hline
\parbox[t][][t]{0.2\textwidth}{\raggedright \textbf{Remove 1 Edge (WS4)}} &  $0.3290$                                           &  $0.0575$                                                         &  $0.6132$                                                           &  $0.3625$ \\ \hline
\parbox[t][][t]{0.2\textwidth}{\raggedright \textbf{Remove 1 Node (WS1)\textcolor{white}{g}}} &  $1.0000$                       &  $0.0000$                                                         &  $0.0000$                                                           &  $0.0001$ \\ \hline 
\parbox[t][][t]{0.2\textwidth}{\raggedright \textbf{Remove 1 Node (WS3)\textcolor{white}{g}}} &  $0.2500$                       &  $0.0000$                                                         &  $0.7500$                                                           &  $0.0003$ \\ \hline 
\parbox[t][][t]{0.2\textwidth}{\raggedright \textbf{Remove 1 Node (WS4)\textcolor{white}{g}}} &  $0.0000$                       &  $1.0000$                                                         &  $0.0000$                                                           &  $0.000001$ \\ \hline \hline
\parbox[t][][t]{0.4\textwidth}{\raggedright \textbf{Eq. Weighted Avg. (WS1)}} &  $0.5507$                                &  $0.0239$                                                         &  $0.4246$                                                           &           \\ \hline 
\parbox[t][][t]{0.4\textwidth}{\raggedright \textbf{Eq. Weighted Avg. (WS3)}} &  $0.3550$                                &  $0.0511$                                                         &  $0.5939$                                                           &           \\ \hline 
\parbox[t][][t]{0.4\textwidth}{\raggedright \textbf{Eq. Weighted Avg. (WS4)}} &  $0.3006$                                &  $0.3046$                                                         &  $0.3947$                                                           &           \\ \hline 
\parbox[t][][t]{0.4\textwidth}{\raggedright \textbf{Avg. Weighted by Dataset Prop. (WS1)}} &  $0.3720$                  &  $0.0321$                                                         &  $0.5959$                                                           &           \\ \hline
\parbox[t][][t]{0.4\textwidth}{\raggedright \textbf{Avg. Weighted by Dataset Prop. (WS3)}} &  $0.3055$                  &  $0.0720$                                                         &  $0.6225$                                                           &           \\ \hline
\parbox[t][][t]{0.4\textwidth}{\raggedright \textbf{Avg. Weighted by Dataset Prop. (WS4)}} &  $0.3563$                  &  $0.0593$                                                         &  $0.5843$                                                           &           \\ \hline
\end{tabular}
\caption[Single changes -- inner shape change enhancements: better, equal, or worse in percent]{Single changes -- inner shape change enhancements: better, equal, or worse in percent results with respect to the relative white space average metric.}
\label{tab:tabelle15}
\end{table}

\begin{table}[H]
	\centering
\begin{tabular}{|l|l|l|l|l|l|l|l|}
\hline
                                       & \parbox[t][][t]{0.12\textwidth}{\raggedright \textbf{Aes-\\thetic Criteria Avg. (Base Implementation)}} & \parbox[t][][t]{0.07\textwidth}{\raggedright \textbf{Aes-\\thetic Criteria Avg. (SCE Layout)}} & \parbox[t][][t]{0.07\textwidth}{\textbf{Aes-\\thetic Criteria Avg. Absolute Difference}} & \parbox[t][][t]{0.07\textwidth}{\textbf{Aes-\\thetic Criteria Avg. Metric Ratio}} & \parbox[t][][t]{0.10\textwidth}{\raggedright \textbf{Edge Crossings Avg. (Base Implementation)}} & \parbox[t][][t]{0.12\textwidth}{\raggedright \textbf{Edge Crossings Avg. (Shape Change Enhancing (SCE) Layout)}} & \parbox[t][][t]{0.15\textwidth}{\textbf{Dataset \\Proportion}} \\ \hline
\parbox[t][][t]{0.19\textwidth}{\raggedright \textbf{Add 1 Edge (WS1)}} &  $0.6763$                                 & $0.6373$                     & $0.0390$                                    & $0.9424$              & $173.1443$              & $181.3176$             & $0.6178$                   \\ \hline
\parbox[t][][t]{0.19\textwidth}{\raggedright \textbf{Add 1 Edge (WS3)}} &  $0.6750$                                 & $0.6349$                     & $0.0400$                                    & $0.9408$              & $172.3088$              & $181.3227$             & $0.6238$                   \\ \hline
\parbox[t][][t]{0.19\textwidth}{\raggedright \textbf{Add 1 Edge (WS4)}} &  $0.6753$                                 & $0.6401$                     & $0.0353$                                    & $0.9478$              & $171.9943$              & $172.3701$             & $0.6166$                   \\ \hline
\parbox[t][][t]{0.19\textwidth}{\raggedright \textbf{Add 1 Node (WS1)\textcolor{white}{g}}} &  $0.6671$             & $0.6195$                     & $0.0476$                                    & $0.9286$              & $173.1667$              & $176.1296$             & $0.0278$                   \\ \hline
\parbox[t][][t]{0.19\textwidth}{\raggedright \textbf{Add 1 Node (WS3)\textcolor{white}{g}}} &  $0.6669$             & $0.6208$                     & $0.0460$                                    & $0.9310$              & $176.6696$              & $181.0352$             & $0.0191$                   \\ \hline
\parbox[t][][t]{0.19\textwidth}{\raggedright \textbf{Add 1 Node (WS4)\textcolor{white}{g}}} &  $0.6638$             & $0.6291$                     & $0.0347$                                    & $0.9478$              & $176.0041$              & $176.4449$             & $0.0209$                   \\ \hline
\parbox[t][][t]{0.19\textwidth}{\raggedright \textbf{Remove 1 Edge (WS1)}} & $0.6727$                                & $0.6306$                     & $0.0421$                                    & $0.9374$              & $156.9583$              & $162.6132$             & $0.3542$                    \\ \hline
\parbox[t][][t]{0.19\textwidth}{\raggedright \textbf{Remove 1 Edge (WS3)}} & $0.6754$                                & $0.6333$                     & $0.0421$                                    & $0.9377$              & $157.4276$              & $164.8217$             & $0.3568$                    \\ \hline
\parbox[t][][t]{0.19\textwidth}{\raggedright \textbf{Remove 1 Edge (WS4)}} & $0.6752$                                & $0.6403$                     & $0.0349$                                    & $0.9483$              & $157.1276$              & $157.3821$             & $0.3625$                    \\ \hline
\parbox[t][][t]{0.19\textwidth}{\raggedright \textbf{Remove 1 Node (WS1)}} & $0.6809$           & $0.6067$                     & $0.0742$                                    & $0.8911$              & $138.5000$              & $150.5000$             & $0.0001$                   \\ \hline 
\parbox[t][][t]{0.19\textwidth}{\raggedright \textbf{Remove 1 Node (WS3)}} & $0.6435$           & $0.5919$                     & $0.0517$                                    & $0.9197$              & $132.2500$              & $139.0000$             & $0.0003$                   \\ \hline 
\parbox[t][][t]{0.19\textwidth}{\raggedright \textbf{Remove 1 Node (WS4)}} & $0.6836$           & $0.6839$                     & $0.0000$                                    & $1.0004$              & $124.0000$              & $124.0000$             & $0.000001$                   \\ \hline \hline
\parbox[t][][t]{0.22\textwidth}{\raggedright \textbf{Eq. Weighted Avg. (WS1)}} & $0.6743$                    & $0.6235$                     & $0.0507$                                    & $0.9249$              & $160.4423$              & $167.6401$             &                            \\ \hline 
\parbox[t][][t]{0.22\textwidth}{\raggedright \textbf{Eq. Weighted Avg. (WS3)}} & $0.6652$                    & $0.6202$                     & $0.0450$                                    & $0.9323$              & $159.6640$              & $166.5449$             &                            \\ \hline 
\parbox[t][][t]{0.22\textwidth}{\raggedright \textbf{Eq. Weighted Avg. (WS4)}} & $0.6745$                    & $0.6483$                     & $0.0262$                                    & $0.9611$              & $157.2815$              & $157.5493$             &                            \\ \hline 
\parbox[t][][t]{0.24\textwidth}{\raggedright \textbf{Avg. Weighted by Dataset (DS) Prop. (WS1)}} & $0.6748$      & $0.6344$                     & $0.0403$                                    & $0.9402$              & $167.4058$              & $174.5428$             &                            \\ \hline
\parbox[t][][t]{0.27\textwidth}{\raggedright \textbf{Avg. Weighted by DS Prop. (WS3)}} & $0.6750$      & $0.6341$                     & $0.0409$                                    & $0.9394$              & $167.0694$              & $175.4157$             &                            \\ \hline
\parbox[t][][t]{0.27\textwidth}{\raggedright \textbf{Avg. Weighted by DS Prop. (WS4)}} & $0.6750$      & $0.6399$                     & $0.0351$                                    & $0.0480$              & $166.6849$              & $167.0181$             &                            \\ \hline
\end{tabular}
\vspace{0.4cm}
\caption[Single changes -- inner shape change enhancements: aesthetic criteria average and edge crossings result]{Single changes -- inner shape change enhancements: aesthetic criteria average and edge crossings result.}
\label{tab:tabelle16}
\end{table}

%
%

\textbf{Relative White Space Average.} 
WS1 achieves relatively, weighted according to dataset proportion, a $2.80$-fold higher value than the base implementation (cf. Table~\ref{tab:tabelle14}). Leading is the WS3 approach with a factor of $3.40$ -- i.e., WS3 produces white spaces that are $3.40$ times as large as those of the base implementation. The worst performing approach is WS4. It generates white spaces that are only $1.17$ times larger than those of the base implementation. As for the inner shape change enhancements, slightly more shape change enhancements are reset, because the base implementation performs better than our layout in about $30\%$ of all alternatives (cf. Table~\ref{tab:tabelle15}). Both are equally good in $3-7\%$ of all alternatives. In about $60\%$ of all alternatives, our layout performs better than the base implementation.

\textbf{Aesthetic Criteria Average.}
According to the weighted average by dataset proportion of the aesthetics criteria, the WS4 approach is the best with a loss of about $5.20\%$ (cf. Table~\ref{tab:tabelle16}). However, this approach also achieves the worst results in terms of white space enlargement. It leads to an enlargement of only $1.17$. Since it preserves the horizontal and vertical node distance ratios, it leads to fewer new edge crossings, see Figure~\ref{fig:pictures_DACH_ShapeLayout_WhiteSpaceApproaches}. However, keeping the distance ratios also leads WS4 to achieve the worst result in white space enlargement around the graph change. Furthermore, we can state that the approach WS1, which performs second best in terms of white space enlargement, leads to only $0.78\%$ more loss than WS4 (cf. Table~\ref{tab:tabelle16}). WS3 again follows close behind with a loss of $6.06\%$ (cf. Table~\ref{tab:tabelle16}). The inner shape change enhancements result in about twice the loss in aesthetic criteria average compared to the outer shape change enhancements. This is still a very good result, as it is still well below the tolerance parameter of $10\%$. Furthermore, the inner shape change enhancements tend to be even more intensive interventions to the graph layout, since more graph elements have to be moved to create white space. Thus, the larger loss can be explained by this. As far as edge crossings are concerned, WS4 is the best (cf. Table~\ref{tab:tabelle16}). It leads to $0.33$ more edge crossings than the base implementation. However, as discussed earlier for the aesthetic criteria average, WS4 creates the least white space enlargement. WS1 and WS3 -- the approaches that do not retain the node distance ratios -- create $7.13$ and $8.35$ more edge crossings according to the dataset proportion weighted average (cf. Table~\ref{tab:tabelle16}). The approaches that do not keep the distance ratios tend to generate more e.g. new edge crossings. This is already evident in Figure~\ref{fig:pictures_DACH_ShapeLayout_WhiteSpaceApproaches}. This tendency can be explained by the fact that some parts of the DAG are shifted more than others. With the underlying graph size, however, this is still a really good result which, in combination with the aesthetic criteria average and the relative white space average, allows the conclusion that, also for the inner shape change enhancements, the optimizations and the tolerance parameters fulfill their purpose.

\textbf{Conclusion.} Our layout can also convince with its results for the inner shape change enhancements: In about $60\%$ of all alternatives our layout outperforms the base implementation. The cost in terms of the aesthetics criteria is $6.1\%$ in the worst case while the allowed loss is $10\%$. This shows that the optimization functions achieve their goal.

\FloatBarrier

\paragraph{Single Changes -- Example Visualization Results.} 
\label{par:single_changes_example_result_images}

\begin{figure}[H]
  \centering
    \includegraphics[width=.9\textwidth]{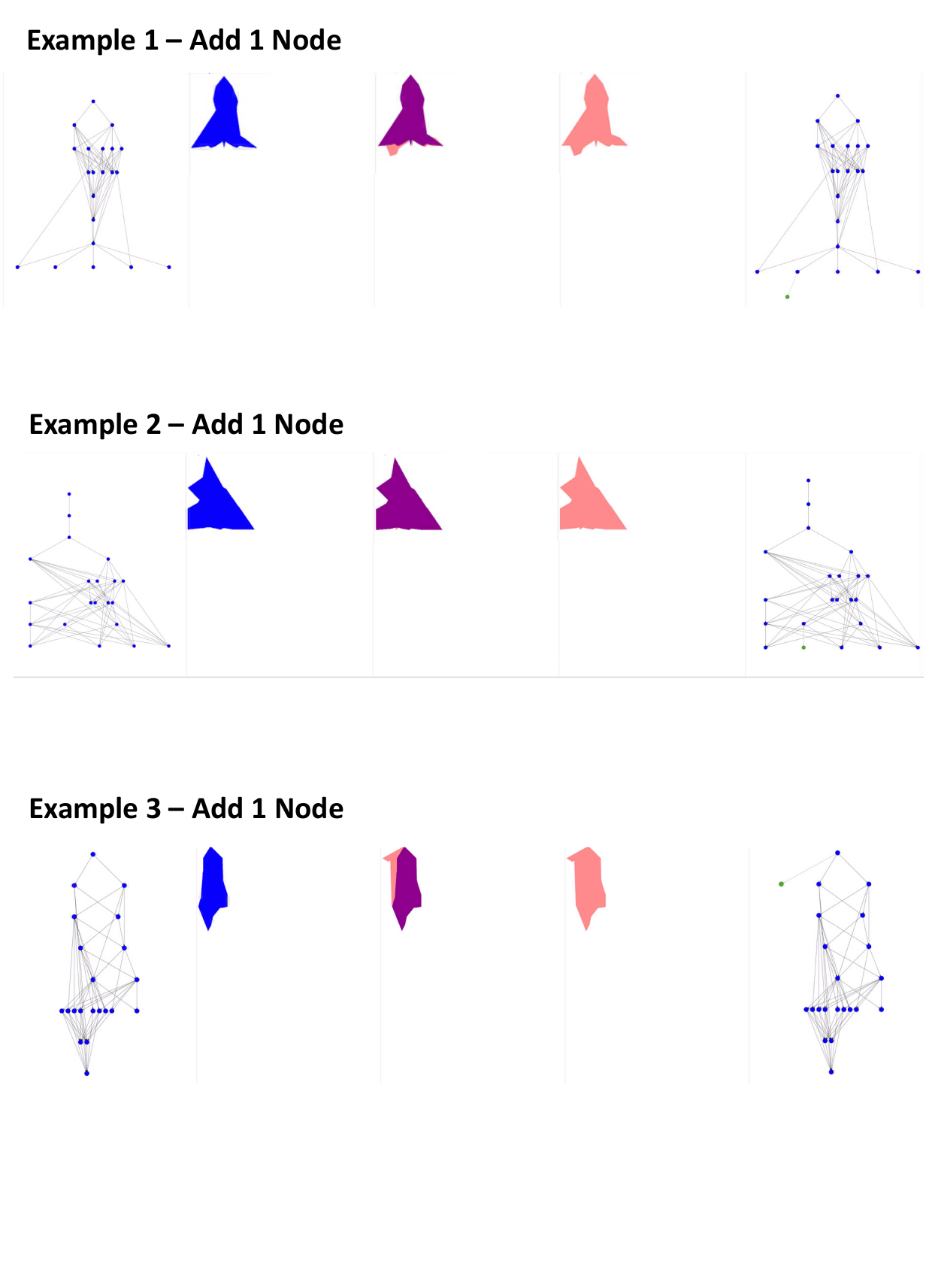}
  \caption[Single graph changes -- example visualization results for single node changes and outer shape change enhancements]{Single graph changes -- example visualization results for single node changes and outer shape change enhancements.}
  \label{fig:pictures_shapeLayout_Graph_Examples_SingleChanges_nodeChanges_outer}
\end{figure}

\begin{figure}[H]
  \centering
    \includegraphics[width=.9\textwidth]{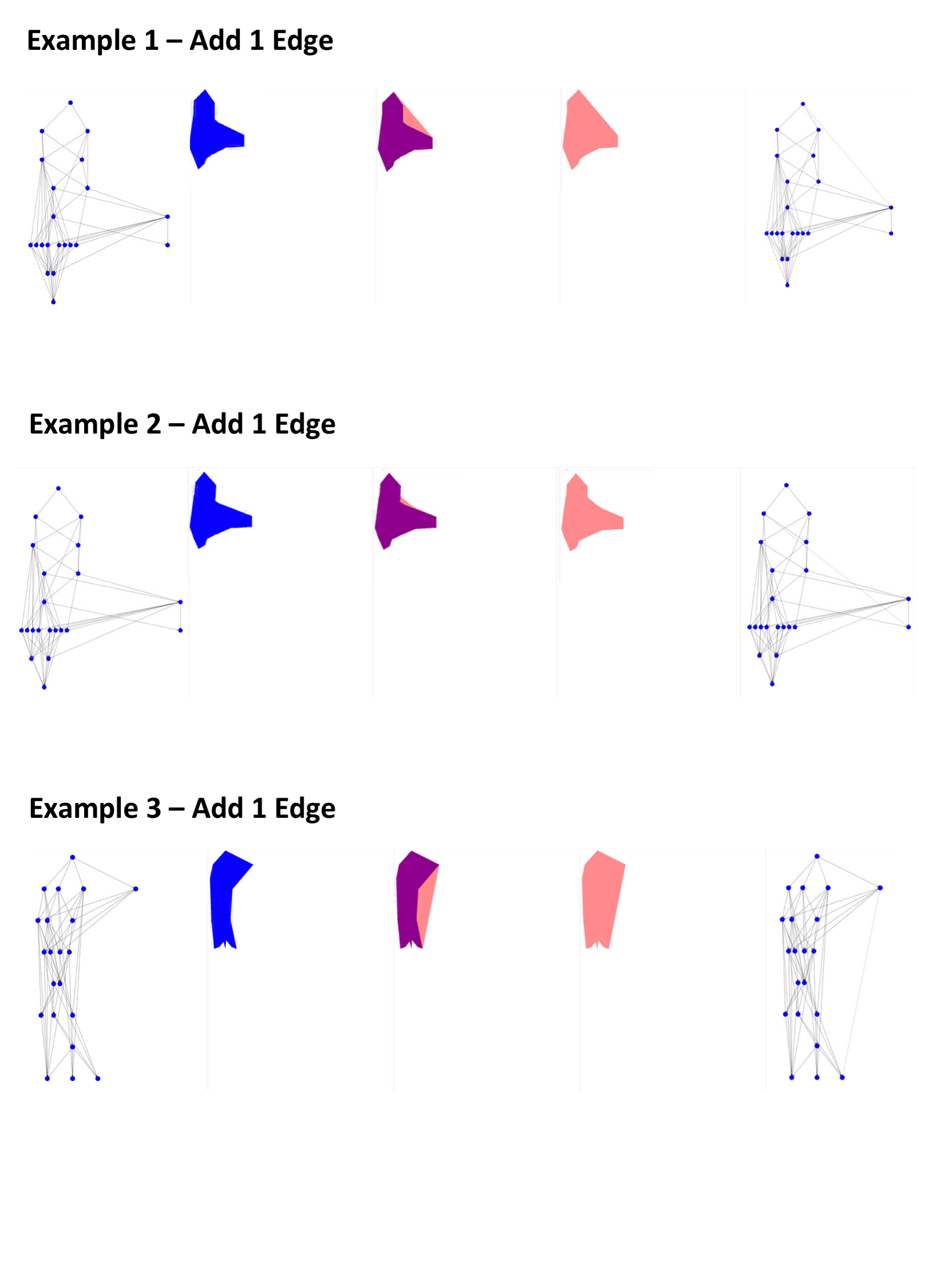}
  \caption[Single graph changes -- example visualization results for single edge changes and outer shape change enhancements]{Single graph changes -- example visualization results for single edge changes and outer shape change enhancements.}
  \label{fig:pictures_shapeLayout_Graph_Examples_SingleChanges_edgeChanges_outer}
\end{figure}

\begin{figure}[H]
  \centering
    \includegraphics[width=.9\textwidth]{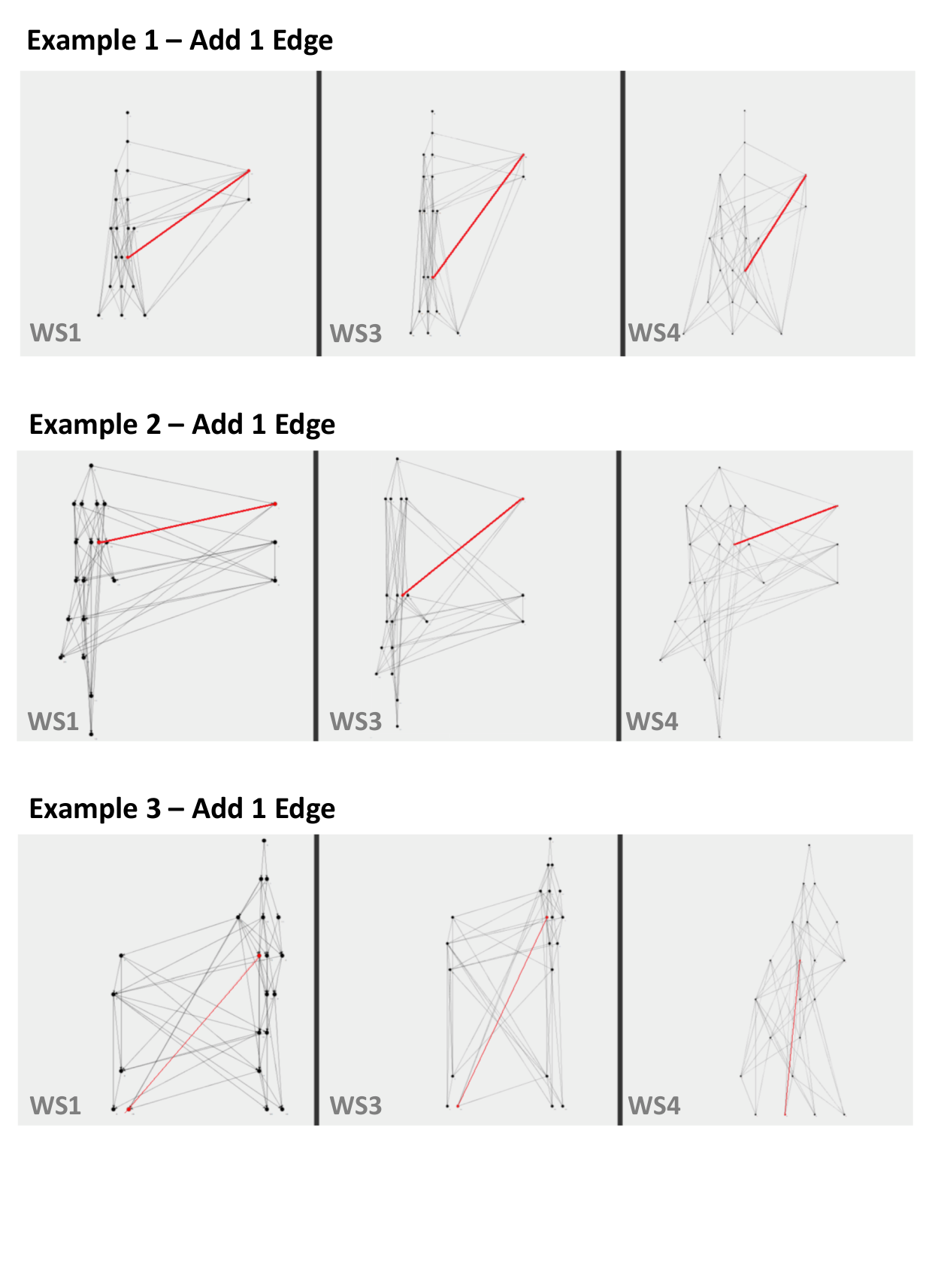}
  \caption[Single graph changes -- example visualization results for single edge changes and inner shape change enhancements]{Single graph changes -- example visualization results for single edge changes and inner shape change enhancements.}
  \label{fig:pictures_shapeLayout_Graph_Examples_SingleChanges_edgeChanges}
\end{figure}

\begin{figure}[H]
  \centering
    \includegraphics[width=.9\textwidth]{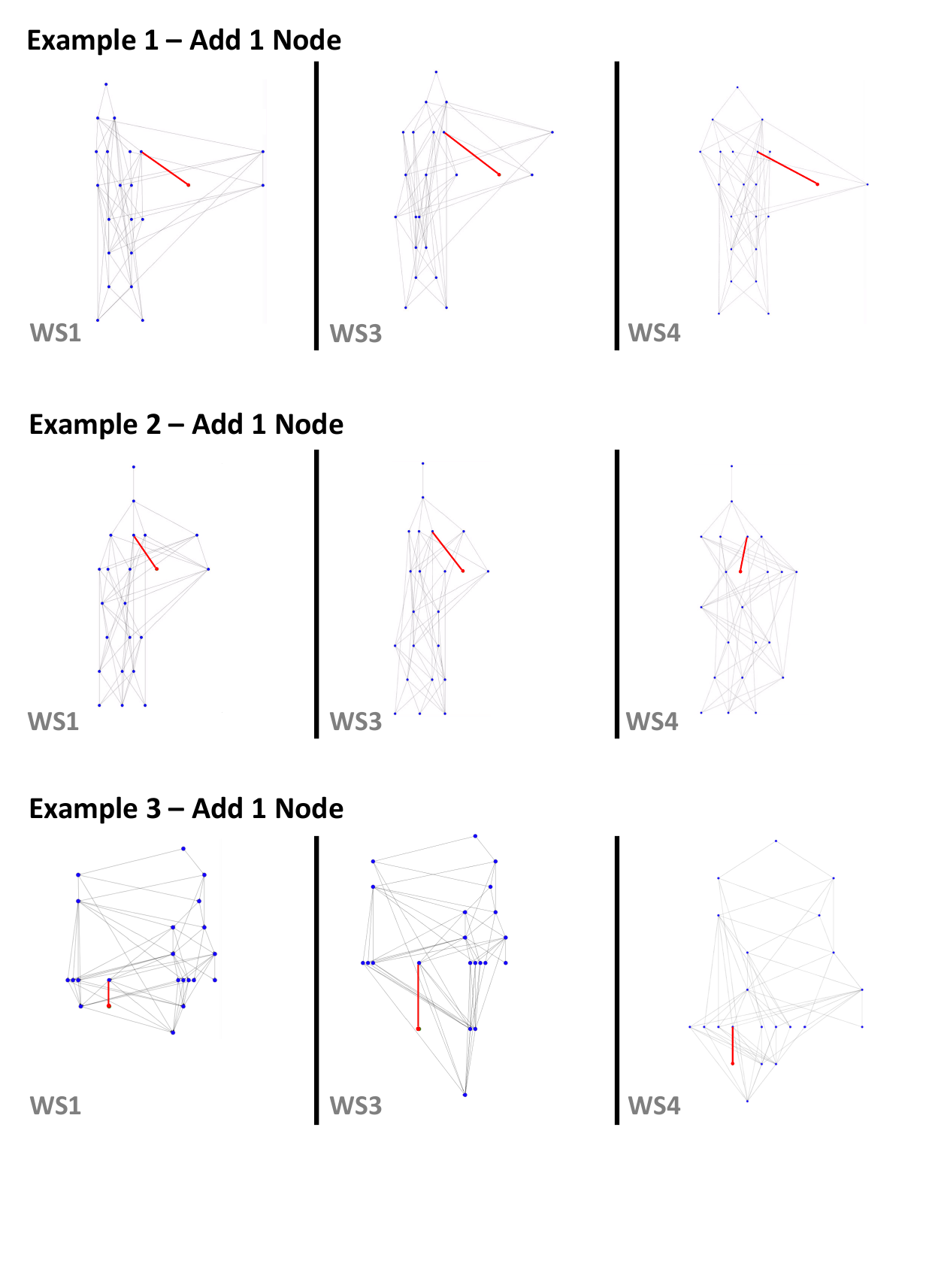}
  \caption[Single graph changes -- example visualization results for single node changes and inner shape change enhancements]{Single graph changes -- example visualization results for single node changes and inner shape change enhancements.}
  \label{fig:pictures_shapeLayout_Graph_Examples_SingleChanges_nodeChanges}
\end{figure}

\textbf{Outer Shape Change Enhancements.} Figure~\ref{fig:pictures_shapeLayout_Graph_Examples_SingleChanges_nodeChanges_outer} shows three visualization results for outer shape change enhancements for adding a node as an example. Example 1 shows the case where the new node is the only node in a layer. Since the node generates more outer hull change when moved to the left compared to the whole graph, the algorithm chooses this option. This is also the reason why the node is not placed directly below its parent node, as is actually typical. Example 2 shows the case where the shape relevant node does not have the minimum or maximum index of a level. This splitting of the graph is similar to the inner shape change enhancements, as it is also based on moving parts of the graph. In this example, the graph split provides a relatively small change to the outer hull, which is due to the path of the edges that are around the new node -- not the optimization itself. The visualized graph shows well that the optimization itself works well. The part of the graph to the left of the new node and the part of the graph to the right of the new node have been moved to the left and to the right, respectively, as desired. Example 3 shows how the outer hull was noticeably changed by an added node that was rotated outward and moved to the left.\\
For the addition of an edge, the same picture is obtained. The optimizations work (cf. Figure~\ref{fig:pictures_shapeLayout_Graph_Examples_SingleChanges_edgeChanges_outer} -- \texttt{Example 1-3}). But as example 1 and 2 clearly show the then resulting change of the outer hull also depends noticeably on the course of the added edge, as well as the further edges. Example 3 shows how a full outer edge in combination with our optimization for shape relevant edges leads to a significant change of the outer hull of the graph (cf. Figure~\ref{fig:pictures_shapeLayout_Graph_Examples_SingleChanges_edgeChanges_outer} -- \texttt{Example 3}). Thus, the visualization examples for outer shape change enhancements support the previous evaluation results.

\textbf{Inner Shape Change Enhancements.} The three examples, each with an edge added (cf. Figure~\ref{fig:pictures_shapeLayout_Graph_Examples_SingleChanges_edgeChanges}), clearly show the better performance of the white space approaches that do not respect the horizontal and vertical node distance ratio. WS1 and WS3 reliably generate white space around the graph changes, which then helps to better detect the inner graph change. WS4 confirms the impression already outlined by Figure~\ref{fig:pictures_DACH_ShapeLayout_WhiteSpaceApproaches}. WS4 enlarges white spaces. But usually not directly at the change where the white space is supposed to be enlarged, but, due to maintaining the distance relation, the white space, which is already larger, is further enlarged. If this is not by chance, as in example 2, the white space that is at the change, another white space area is enlarged, but this one does not contain any change.
For adding single nodes, cf. Figure~\ref{fig:pictures_shapeLayout_Graph_Examples_SingleChanges_nodeChanges}, the same pattern emerges. The white space approaches WS1 and WS3 reliably create enlarged white space around the inner graph change while the WS4 approach again tends to increase white spaces that are already larger. If the already larger white space does not happen to be at the change, as in example 2, another white space area is enlarged that does not contain a change.

\textbf{Conclusion.} In graph drawing one always tries to draw the graphs as symmetrical as possible, because it is well known that people find symmetrical visualizations pleasant or even beautiful \cite{reber2002reasons}. But as it becomes clear here, for the enlargement of certain white spaces -- here: the white spaces at the graph changes -- a certain asymmetry is necessary; i.e. the non-compliance with the horizontal and vertical node distance ratio. WS1 and WS3 generate white space enlargements around graph changes much more reliably, but, for example, the level distances are no longer uniform across all hierarchical levels of the graph (WS3) (cf. Figure~\ref{fig:pictures_DACH_ShapeLayout_WhiteSpaceApproaches}). This is a certain break in the drawing symmetry of the graph. The same is to be noted for the outer shape change enhancements. Also the graph splitting and the left/right moving of nodes cause a certain asymmetry and therefore a certain break of the drawing symmetry.
\FloatBarrier

\paragraph{Multiple Simultaneous Changes} 
\label{par:multiple_simultaneous_changes_outer_shape_change_enhancement_evaluation_results}

\begin{table}[H]
	\centering
\begin{tabular}{|l|l|l|l|l|l|}
\hline
                                       & \parbox[t][][t]{0.15\textwidth}{\raggedright \textbf{Intersection over Union (Base Implementation)}} & \parbox[t][][t]{0.15\textwidth}{\raggedright \textbf{Intersection over Union (Shape Change Enhancing Layout)}} & \parbox[t][][t]{0.12\textwidth}{\textbf{Absolute Difference}} & \parbox[t][][t]{0.12\textwidth}{\textbf{Metric Ratio}} & \parbox[t][][t]{0.15\textwidth}{\textbf{Dataset \\Proportion}} \\ \hline
\parbox[t][][t]{0.2\textwidth}{\raggedright \textbf{Add 3 Nodes (WS1)}} &  $0.1325$                                              & $0.2688$                                    & $0.1363$                               & $2.0284$              & $0.5965$                    \\ \hline
\parbox[t][][t]{0.2\textwidth}{\raggedright \textbf{Add 3 Nodes (WS3)}} &  $0.1394$                                              & $0.2832$                                    & $0.1438$                               & $2.0313$              & $0.5967$                    \\ \hline
\parbox[t][][t]{0.2\textwidth}{\raggedright \textbf{Add 3 Nodes (WS4)}} &  $0.1394$                                              & $0.2940$                                    & $0.1545$                               & $2.1086$              & $0.5967$                    \\ \hline
\parbox[t][][t]{0.2\textwidth}{\raggedright \textbf{Add 1 Node, Add 2 Edges (WS1)}} &  $0.0584$                                  & $0.0828$                                    & $0.0244$                               & $1.4170$              & $0.4035$                    \\ \hline
\parbox[t][][t]{0.2\textwidth}{\raggedright \textbf{Add 1 Node, Add 2 Edges (WS3)}} &  $0.0575$                                  & $0.0810$                                    & $0.0236$                               & $1.4098$              & $0.4033$                    \\ \hline
\parbox[t][][t]{0.2\textwidth}{\raggedright \textbf{Add 1 Node, Add 2 Edges (WS4)}} &  $0.0590$                                  & $0.1437$                                    & $0.0847$                               & $2.4372$              & $0.4033$                    \\ \hline \hline
\parbox[t][][t]{0.2\textwidth}{\raggedright \textbf{Eq. Weighted Avg. (WS1)}} & $0.0955$                                  & $0.1758$                                    & $0.0803$                               & $1.7227$              &                                        \\ \hline
\parbox[t][][t]{0.2\textwidth}{\raggedright \textbf{Eq. Weighted Avg. (WS3)}} & $0.0985$                                  & $0.1821$                                    & $0.0837$                               & $1.7295$              &                                        \\ \hline
\parbox[t][][t]{0.2\textwidth}{\raggedright \textbf{Eq. Weighted Avg. (WS4)}} & $0.0992$                                  & $0.2188$                                    & $0.1196$                               & $2.2729$              &                                        \\ \hline
\parbox[t][][t]{0.2\textwidth}{\raggedright \textbf{Avg. Weighted by Dataset Prop. (WS1)}} & $0.1026$                    & $0.1937$                                    & $0.0911$                               & $1.7817$              &                                        \\ \hline
\parbox[t][][t]{0.2\textwidth}{\raggedright \textbf{Avg. Weighted by Dataset Prop. (WS3)}} & $0.1064$                    & $0.2017$                                    & $0.0953$                               & $1.7806$              &                                        \\ \hline
\parbox[t][][t]{0.2\textwidth}{\raggedright \textbf{Avg. Weighted by Dataset Prop. (WS4)}} & $0.1070$                    & $0.2334$                                    & $0.1264$                               & $2.2411$              &                                        \\ \hline
\end{tabular}
\caption[Multiple simultaneous changes -- outer shape change enhancements: intersection over union results]{Multiple simultaneous changes -- outer shape change enhancements: intersection over union results.}
\label{tab:tabelle19a}
\end{table}

\begin{table}[H]
	\centering
\begin{tabular}{|l|l|l|l|l|}
\hline
                                       & \parbox[t][][t]{0.15\textwidth}{\raggedright \textbf{Base Implementation Better}} & \parbox[t][][t]{0.15\textwidth}{\raggedright \textbf{Base Implementation = Shape Change Enhancing Layout}} & \parbox[t][][t]{0.12\textwidth}{\textbf{Shape Change Enhancing Layout Better}} & \parbox[t][][t]{0.15\textwidth}{\raggedright \textbf{Dataset \\Proportion}} \\ \hline
\parbox[t][][t]{0.2\textwidth}{\raggedright \textbf{Add 3 Nodes (WS1)}} &  $0.0097$                                             &  $0.0147 $                                                        &  $0.9755$                                                           &  $0.5965$ \\ \hline
\parbox[t][][t]{0.2\textwidth}{\raggedright \textbf{Add 3 Nodes (WS3)}} &  $0.0138$                                             &  $0.0095$                                                         &  $0.9767$                                                           &  $0.5967$ \\ \hline
\parbox[t][][t]{0.2\textwidth}{\raggedright \textbf{Add 3 Nodes (WS4)}} &  $0.0051$                                             &  $0.0119$                                                         &  $0.9831$                                                           &  $0.5967$ \\ \hline
\parbox[t][][t]{0.2\textwidth}{\raggedright \textbf{Add 1 Node, Add 2 Edges (WS1)}} &  $0.0361 $                                &  $0.1095$                                                         &  $0.8544$                                                           &  $0.4035$ \\ \hline
\parbox[t][][t]{0.2\textwidth}{\raggedright \textbf{Add 1 Node, Add 2 Edges (WS3)}} &  $0.1405$                                 &  $0.1167$                                                         &  $0.7428$                                                           &  $0.4033$ \\ \hline
\parbox[t][][t]{0.2\textwidth}{\raggedright \textbf{Add 1 Node, Add 2 Edges (WS4)}} &  $0.0299$                                 &  $0.1046$                                                         &  $0.8655$                                                           &  $0.4033$ \\ \hline \hline
\parbox[t][][t]{0.2\textwidth}{\raggedright \textbf{Eq. Weighted Avg. (WS1)}} &  $0.0229$                                &  $0.0621$                                                         &  $0.9149$                                                           &           \\ \hline 
\parbox[t][][t]{0.2\textwidth}{\raggedright \textbf{Eq. Weighted Avg. (WS3)}} &  $0.0772$                                &  $0.0631$                                                         &  $0.8597$                                                           &           \\ \hline 
\parbox[t][][t]{0.2\textwidth}{\raggedright \textbf{Eq. Weighted Avg. (WS4)}} &  $0.0175$                                &  $0.0583$                                                         &  $0.9243$                                                           &           \\ \hline 
\parbox[t][][t]{0.2\textwidth}{\raggedright \textbf{Avg. Weighted by Dataset Prop. (WS1)}} &  $0.0203$                  &  $0.0530$                                                         &  $0.9266$                                                           &           \\ \hline
\parbox[t][][t]{0.2\textwidth}{\raggedright \textbf{Avg. Weighted by Dataset Prop. (WS3)}} &  $0.0649$                  &  $0.0527$                                                         &  $0.8823$                                                           &           \\ \hline
\parbox[t][][t]{0.2\textwidth}{\raggedright \textbf{Avg. Weighted by Dataset Prop. (WS4)}} &  $0.0151$                  &  $0.0493$                                                         &  $0.9356$                                                           &           \\ \hline
\end{tabular}
\caption[Multiple simultaneous changes -- outer shape change enhancements: better, equal, or worse in percent results]{Multiple simultaneous changes -- outer shape change enhancements: better, equal, or worse in percent results with respect to the intersection over union metric.}
\label{tab:tabelle19b}
\end{table}

\begin{table}[H]
	\centering
\begin{tabular}{|l|l|l|l|l|l|}
\hline
                                       & \parbox[t][][t]{0.15\textwidth}{\raggedright \textbf{Relative White Space Average (Base Implementation)}} & \parbox[t][][t]{0.15\textwidth}{\raggedright \textbf{Relative White Space Average (Shape Change Enhancing Layout)}} & \parbox[t][][t]{0.12\textwidth}{\textbf{Absolute Difference}} & \parbox[t][][t]{0.12\textwidth}{\textbf{Metric Ratio}} & \parbox[t][][t]{0.15\textwidth}{\textbf{Dataset \\Proportion}} \\ \hline
\parbox[t][][t]{0.2\textwidth}{\raggedright \textbf{Add 3 Nodes (WS1)}} &  $0.0020$                                             &  $0.0042$                                                         &  $0.0018$                       & $2.0417$                                     &  $0.5965$ \\ \hline
\parbox[t][][t]{0.2\textwidth}{\raggedright \textbf{Add 3 Nodes (WS3)}} &  $0.0018$                                             &  $0.0043$                                                         &  $0.0022$                       & $2.3533$                                     &  $0.5967$ \\ \hline
\parbox[t][][t]{0.2\textwidth}{\raggedright \textbf{Add 3 Nodes (WS4)}} &  $0.0019$                                             &  $0.0018$                                                         &  $0.000001$                     & $0.9482$                                     &  $0.5967$ \\ \hline
\parbox[t][][t]{0.2\textwidth}{\raggedright \textbf{Add 1 Node, Add 2 Edges (WS1)}} &  $0.0013$                                 &  $0.0051$                                                         &  $0.0035$                       & $3.8308$                                     &  $0.4035$ \\ \hline
\parbox[t][][t]{0.2\textwidth}{\raggedright \textbf{Add 1 Node, Add 2 Edges (WS3)}} &  $0.0013$                                 &  $0.0061$                                                         &  $0.0044$                       & $4.6106$                                     &  $0.4033$ \\ \hline
\parbox[t][][t]{0.2\textwidth}{\raggedright \textbf{Add 1 Node, Add 2 Edges (WS4)}} &  $0.0014$                                 &  $0.0016$                                                         &  $0.0002$                       & $1.1944$                                     &  $0.4033$ \\ \hline \hline
\parbox[t][][t]{0.2\textwidth}{\raggedright \textbf{Eq. Weighted Avg. (WS1)}} &  $0.0017$                                &  $0.0046$                                                         &  $0.0026$                       & $2.9362$                                     &           \\ \hline 
\parbox[t][][t]{0.2\textwidth}{\raggedright \textbf{Eq. Weighted Avg. (WS3)}} &  $0.0016$                                &  $0.0052$                                                         &  $0.0033$                       & $3.4819$                                     &           \\ \hline 
\parbox[t][][t]{0.2\textwidth}{\raggedright \textbf{Eq. Weighted Avg. (WS4)}} &  $0.0016$                                &  $0.0017$                                                         &  $0.0001$                       & $1.0713$                                     &           \\ \hline 
\parbox[t][][t]{0.2\textwidth}{\raggedright \textbf{Avg. Weighted by Dataset Prop. (WS1)}} &  $0.0018$                  &  $0.0045$                                                         &  $0.0025$                       & $2.7636$                                     &           \\ \hline
\parbox[t][][t]{0.2\textwidth}{\raggedright \textbf{Avg. Weighted by Dataset Prop. (WS3)}} &  $0.0016$                  &  $0.0050$                                                         &  $0.0030$                       & $3.2637$                                     &           \\ \hline
\parbox[t][][t]{0.2\textwidth}{\raggedright \textbf{Avg. Weighted by Dataset Prop. (WS4)}} &  $0.0017$                  &  $0.0017$                                                         &  $0.0001$                       & $1.0475$                                     &           \\ \hline
\end{tabular}
\caption[Multiple simultaneous changes -- inner shape change enhancements: relative white space average results]{Multiple simultaneous changes -- inner shape change enhancements: relative white space average results.}
\label{tab:tabelle21a}
\end{table}

\begin{table}[H]
	\centering
\begin{tabular}{|l|l|l|l|l|}
\hline
                                       & \parbox[t][][t]{0.15\textwidth}{\raggedright \textbf{Base Implementation Better}} & \parbox[t][][t]{0.15\textwidth}{\raggedright \textbf{Base Implementation = Shape Change Enhancing Layout}} & \parbox[t][][t]{0.12\textwidth}{\textbf{Shape Change Enhancing Layout Better}} & \parbox[t][][t]{0.15\textwidth}{\raggedright \textbf{Dataset \\Proportion}} \\ \hline
\parbox[t][][t]{0.2\textwidth}{\raggedright \textbf{Add 3 Nodes (WS1)}} &  $0.4244$                                              & $0.0009$                                    & $0.5746$                              & $0.5965$                    \\ \hline
\parbox[t][][t]{0.2\textwidth}{\raggedright \textbf{Add 3 Nodes (WS3)}} &  $0.4572$                                              & $0.0034$                                    & $0.5394$                              & $0.5967$                    \\ \hline
\parbox[t][][t]{0.2\textwidth}{\raggedright \textbf{Add 3 Nodes (WS4)}} &  $0.5577$                                              & $0.0064$                                    & $0.4360$                              & $0.5967$                    \\ \hline
\parbox[t][][t]{0.2\textwidth}{\raggedright \textbf{Add 1 Node, Add 2 Edges (WS1)}} &  $0.2933$                                  & $0.0043$                                    & $0.7025$                              & $0.4035$                    \\ \hline
\parbox[t][][t]{0.2\textwidth}{\raggedright \textbf{Add 1 Node, Add 2 Edges (WS3)}} &  $0.3161$                                  & $0.0239$                                    & $0.6600$                              & $0.4033$                    \\ \hline
\parbox[t][][t]{0.2\textwidth}{\raggedright \textbf{Add 1 Node, Add 2 Edges (WS4)}} &  $0.4202$                                  & $0.0116$                                    & $0.5682$                              & $0.4033$                    \\ \hline \hline
\parbox[t][][t]{0.2\textwidth}{\raggedright \textbf{Eq. Weighted Avg. (WS1)}} & $0.3589 $                                 & $0.0026$                                    & $0.6385$                              &                                        \\ \hline
\parbox[t][][t]{0.2\textwidth}{\raggedright \textbf{Eq. Weighted Avg. (WS3)}} & $0.3866$                                  & $0.0137$                                    & $0.5997$                              &                                        \\ \hline
\parbox[t][][t]{0.2\textwidth}{\raggedright \textbf{Eq. Weighted Avg. (WS4)}} & $0.4890$                                  & $0.0090$                                    & $0.5021$                              &                                        \\ \hline
\parbox[t][][t]{0.2\textwidth}{\raggedright \textbf{Avg. Weighted by Dataset Prop. (WS1)}} & $0.3715$                    & $0.0023$                                    & $0.6262$                              &                                        \\ \hline
\parbox[t][][t]{0.2\textwidth}{\raggedright \textbf{Avg. Weighted by Dataset Prop. (WS3)}} & $0.4002$                    & $0.0117$                                    & $0.5881$                              &                                        \\ \hline
\parbox[t][][t]{0.2\textwidth}{\raggedright \textbf{Avg. Weighted by Dataset Prop. (WS4)}} & $0.5022$                    & $0.0085$                                    & $0.4893$                              &                                        \\ \hline
\end{tabular}
\caption[Multiple simultaneous changes -- inner shape change enhancements: better, equal, or worse in percent results]{Multiple simultaneous changes -- inner shape change enhancements: better, equal, or worse in percent results with respect to the relative white space average metric.}
\label{tab:tabelle21b}
\end{table}

\begin{table}[H]
	\centering
\begin{tabular}{|l|l|l|l|l|l|l|l|}
\hline
                                       & \parbox[t][][t]{0.09\textwidth}{\raggedright \textbf{Aes-\\thetic Criteria Average (Base Implementation)}} & \parbox[t][][t]{0.09\textwidth}{\raggedright \textbf{Aes-\\thetic Criteria Average (Shape Change Enhancing Layout)}} & \parbox[t][][t]{0.085\textwidth}{\textbf{Aes-\\thetic Criteria Average Absolute Difference}} & \parbox[t][][t]{0.085\textwidth}{\textbf{Aes-\\thetic Criteria Average Metric Ratio}} & \parbox[t][][t]{0.08\textwidth}{\raggedright \textbf{Edge Crossings Average (Base Implementation)}} & \parbox[t][][t]{0.08\textwidth}{\raggedright \textbf{Edge Crossings Average (Shape Change Enhancing Layout)}} & \parbox[t][][t]{0.10\textwidth}{\textbf{Dataset \\Propor-tion}} \\ \hline
\parbox[t][][t]{0.19\textwidth}{\raggedright \textbf{Add 3 Nodes (WS1)}} &  $0.6768$                                & $0.6304$                     & $0.0464$                                    & $0.9315$              & $153.8489$              & $154.5513$             & $0.5965$                   \\ \hline
\parbox[t][][t]{0.19\textwidth}{\raggedright \textbf{Add 3 Nodes (WS3)}} &  $0.6768$                                & $0.6304$                     & $0.0464$                                    & $0.9315$              & $153.4370$              & $154.7107$             & $0.5967$                   \\ \hline
\parbox[t][][t]{0.19\textwidth}{\raggedright \textbf{Add 3 Nodes (WS4)}} &  $0.6776$                                & $0.6349$                     & $0.0427$                                    & $0.9369$              & $153.0312$              & $153.5835$             & $0.5967$                   \\ \hline
\parbox[t][][t]{0.19\textwidth}{\raggedright \textbf{Add 1 Node, Add 2 Edges (WS1)}} &  $0.6939$                    & $0.6436$                     & $0.0494$                                    & $0.9288$              & $136.8921$              & $142.0985$             & $0.4035$                   \\ \hline
\parbox[t][][t]{0.19\textwidth}{\raggedright \textbf{Add 1 Node, Add 2 Edges (WS3)}} &  $0.6934$                    & $0.6371$                     & $0.0564$                                    & $0.9187$              & $136.8641$              & $141.9567$             & $0.4033$                   \\ \hline
\parbox[t][][t]{0.19\textwidth}{\raggedright \textbf{Add 1 Node, Add 2 Edges (WS4)}} &  $0.6933$                    & $0.6330$                     & $0.0603$                                    & $0.9131$              & $137.2090$              & $136.8804$             & $0.4033$                   \\ \hline \hline
\parbox[t][][t]{0.24\textwidth}{\raggedright \textbf{Eq. Weighted Avg. (WS1)}} & $0.6849$                    & $0.6370$                     & $0.0479$                                    & $0.9301$              & $145.3705$              & $148.3249$             &                            \\ \hline 
\parbox[t][][t]{0.24\textwidth}{\raggedright \textbf{Eq. Weighted Avg. (WS3)}} & $0.6851$                    & $0.6337$                     & $0.0514$                                    & $0.9251$              & $145.1505$              & $148.3337$             &                            \\ \hline 
\parbox[t][][t]{0.24\textwidth}{\raggedright \textbf{Eq. Weighted Avg. (WS4)}} & $0.6854$                    & $0.6339$                     & $0.0515$                                    & $0.9250$              & $145.1201$              & $145.2320$             &                            \\ \hline 
\parbox[t][][t]{0.27\textwidth}{\raggedright \textbf{Avg. Weighted by Dataset Prop. (WS1)}} & $0.6833$      & $0.6357$                     & $0.0476$                                    & $0.9304$              & $147.0070$              & $149.5268$             &                            \\ \hline
\parbox[t][][t]{0.27\textwidth}{\raggedright \textbf{Avg. Weighted by Dataset Prop. (WS3)}} & $0.6835$      & $0.6331$                     & $0.0504$                                    & $0.9263$              & $146.7526$              & $149.5666$             &                            \\ \hline
\parbox[t][][t]{0.27\textwidth}{\raggedright \textbf{Avg. Weighted by Dataset Prop. (WS4)}} & $0.6839$      & $0.6341$                     & $0.0498$                                    & $0.9273$              & $146.6504$              & $146.8475$             &                            \\ \hline
\end{tabular}
\caption[Multiple simultaneous changes: aesthetic criteria average and edge crossings result]{Multiple simultaneous changes: aesthetic criteria average and edge crossings result.}
\label{tab:tabelle22}
\end{table}


\textbf{Outer Shape Change Enhancement Evaluation Results.} 
\textbf{\emph{Intersection over Union.}} For multiple graph changes, using the white space approaches WS1 and WS3, the outer hull changes by $1.78$ times compared to the base implementation (cf. Table~\ref{tab:tabelle19a}). If the white space approach WS4 is used, the outer hull changes even by $2.24$-fold compared to the base implementation (cf. Table~\ref{tab:tabelle19a}). However, here again comes the problem that by maintaining the distance ratios, the WS4 method further enlarges white spaces that are already larger, and thus the white spaces that contain a change are not necessarily the ones that are conspicuously enlarged -- unless the change happens to be in the white space that is already larger and is further enlarged by WS4. We can state that our shape change enhancing layout performs about two times better than the base implementation for multiple simultaneous changes according to the weighted average by dataset proportion (cf. Table~\ref{tab:tabelle19a}). The other evaluations -- equal-weighted average, average per changes, and white space method -- also paint a similar picture. Comparing the proportions of how often our shape change enhancing layout achieves a better result than the base implementation shows that when the white space approaches WS1 and WS3 are used, our shape change enhancing layout achieves a better result than the base implementation in $73.70\%$ and $75.51\%$ of the cases, respectively (cf. Table~\ref{tab:tabelle19b}). If the WS 4 approach is used, our shape change enhancing layout even achieves a better result than the building implementation in $89.93\%$ of the cases. However, it must be emphasized again that the WS4 approach does not increase the desired white space. This of course relativizes the apparently good performance using WS4 again.
\textbf{Inner Shape Change Enhancement Evaluation Results.} 
\textbf{\emph{Relative White Space Average.}} Relative to the base implementation, our WS3 approach again achieves the best result. WS3 achieves a relative white space $3.26$ times as large as that of the base implementation (cf. Table~\ref{tab:tabelle21a}). Our approach WS1 also achieves an excellent result with an average of $2.76$ times as large relative white space (cf. Table~\ref{tab:tabelle21a}). Compared to the base implementation, our approach WS4 achieves on average only a slightly better result than the base implementation itself. WS4 achieves on average a $1.04$ larger relative white space than the base implementation (cf. Table~\ref{tab:tabelle21a}). The results are nearly identical to the inner shape change enhancement results for single changes. Our approach WS1 and WS3 are better than the base implementation in more than and just about $60\%$ of the cases, respectively, in the by dataset proportion weighted average (cf. Table~\ref{tab:tabelle21b}). For the WS4 approach, it is more of a coincidence whether it or the base implementation is better: WS4 is better than the base implementation in $48.92\%$ of the cases and the base implementation is better than our WS4 approach in $50.22\%$ of the cases (cf. Table~\ref{tab:tabelle21b}). In conclusion, also this metric supports the previous evaluation results.

\textbf{Aesthetic Criteria Average.} 
As Table~\ref{tab:tabelle22} shows, for multiple simultaneous graph changes, all three white space approaches have similar costs in terms of aesthetic criteria average. All three approaches result in a loss of about $7\%$. Given the nevertheless very strong interventions -- outer and inner shape change enhancements -- this is a very good result in combination with the shape-relevant adaptations. The results of the multiple simultaneous graph changes for the achieved aesthetic criteria average are comparable to the aesthetic criteria average for the inner shape change enhancements of the single changes. That means the inner shape change enhancements have a stronger influence on the aesthetic criteria average than the outer shape change enhancements (approx. twice (cf. Sections \texttt{Single Changes - Inner Shape Change Enhancement Evaluation Results, Discussion, and Conclusion} -- \texttt{Aesthetic Criteria Average} and \texttt{Single Changes - Outer Shape Change Enhancement Evaluation Results, Discussion, and Conclusion} -- \\\texttt{Aesthetic Criteria Average})), but the influence of the inner shape change enhancements does not become noticeably stronger as the number of graph changes increases.
As far as edge crossings are concerned, WS4 is the best (cf. Table~\ref{tab:tabelle22}). It leads to $0.20$ more edge crossings than the base implementation. However, as already discussed, WS4 creates the least white space enlargement. WS1 and WS3 create $2.52$ and $2.81$ more edge crossings according to the average weighted by dataset proportion (cf. Table~\ref{tab:tabelle22}). The approaches that do not keep the distance ratios tend to generate more e.g. new edge crossings. This is already evident in Figure~\ref{fig:pictures_DACH_ShapeLayout_WhiteSpaceApproaches}. This tendency can be explained by the fact that some parts of the DAG are shifted more than others. With the underlying graph size, however, this is still a really good result which, in combination with the aesthetic criteria average, the intersection over union results, and the relative white space average, allows the conclusion that, also for multiple graph changes, the optimizations and the tolerance parameters fulfill their purpose.

\textbf{Conclusion.} Also for multiple, simultaneously occurring graph changes our shape change enhancing layout achieves convincing results: It outperforms the base implementation in about $75\%$ of all cases when it comes to the intersection over union metric. For the relative white space average metric, our layout outperforms the base implementation in about $60\%$ of all alternatives. The cost in terms of the aesthetics criteria are approximately $7\%$. This shows that the optimization functions achieve their goal.
\FloatBarrier

\paragraph{Multiple Simultaneous Changes -- Example Visualization Results} 
\label{par:multiple_simultaneous_changes_example_visualization_results}
Also the visualization results of multiple simultaneous changes are in line with the afore discussed evaluation results. They show the afore discussed advantages and issues which were already apparent in the visualizations of single graph changes. 

\includepdf[pages=-,frame,pagecommand={},width=0.9\textwidth]{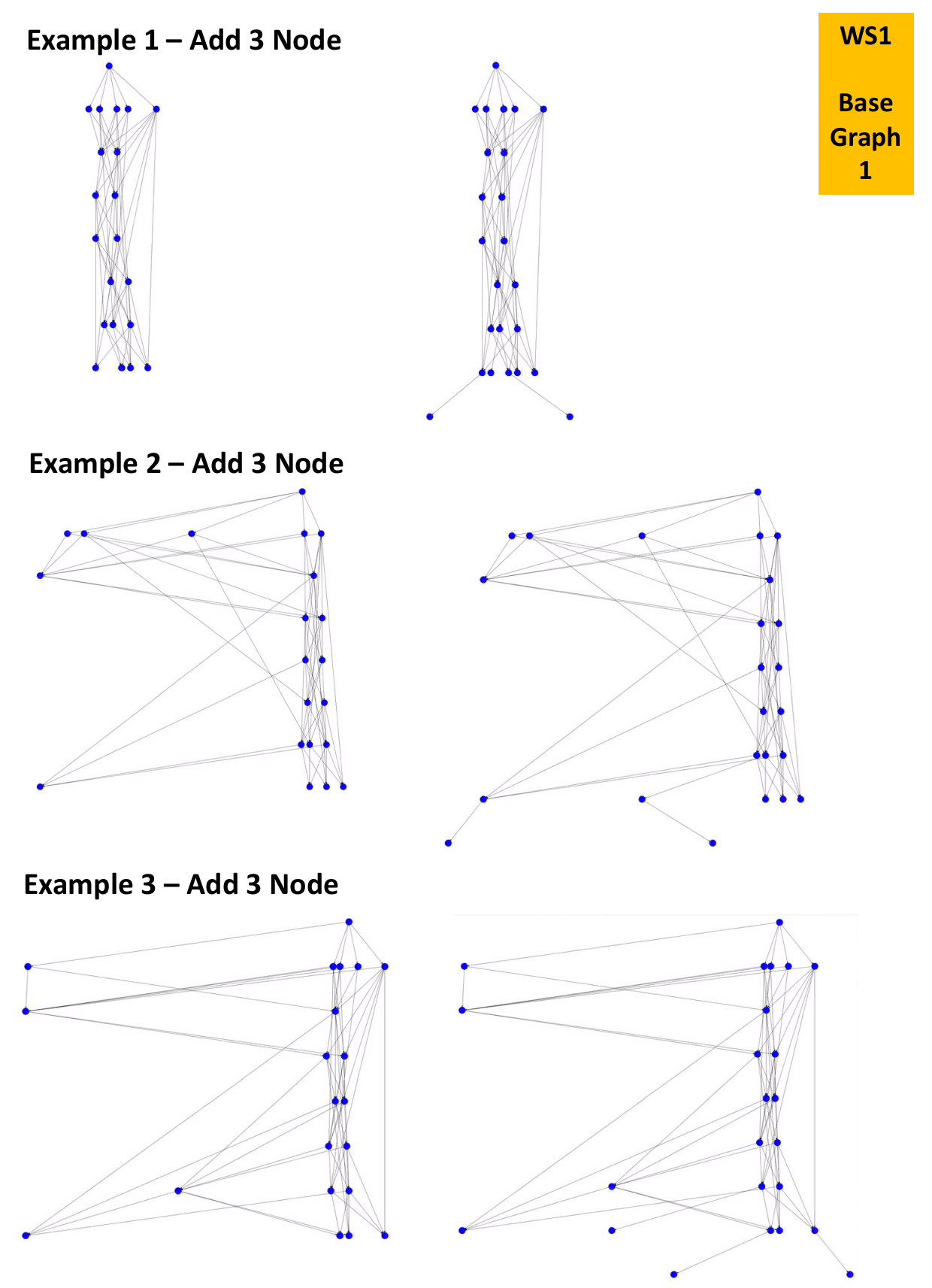}

\subsection{Conclusion, and Future Work} 
\label{sub:discussion_conclusion_and_future_work}
Given the results, we can draw the final conclusion that our newly introduced drawing criteria for
\begin{itemize}
	\item outwardly swapping as many graph changes as possible
	\item changing the outer hull by repositioning the outwardly swapped DAG changes
	\item dealing with DAG changes which are not possible to outwardly swap -- nevertheless we also need a change of the DAG's shape for those as well
\end{itemize}
are successful (keyword: percentage of outperforming the base implementation) and in spite of their intensive interventions in the drawing of the DAGs the resulting drawn DAGs still achieve more than $90\%$ of the aesthetic criteria average compared to the base implementation. This means that our new criteria are intensive interventions and they break with the drawing conventions to some extent but they are still within the conventions which determine whether a human would like a visualization and find it aesthetically pleasing -- otherwise the aesthetic criteria average would not be that good. Because of that and due to the feedback of our participants we can deduce that humans can work with our visualized DAGs and profit from the easier detectability of DAG changes.

In the future it would be interesting to further fine tune our layout's parameters with respect to human perception. This would enable humans to spot differences in pairwise comparisons even better. This requires multiple user studies with an extensive number of participants making it a separate future work topic.\\
Also the positioning of the change within the enhanced white space is an improvement point in the future. If we were able to steer the position of the inner graph change into the center of the white space the chances of the graph change being spotted by a human would increase (our hypothesis). To achieve this  an approach for balancing the size of the minimum (\yellowMinCircle{$\mathtt{Circle_{Min}}$}) and the maximum (\blueMaxCircle{$\mathtt{Circle_{Max}}$}) circle adjacent to the graph change is necessary (cf. Figure~\ref{fig:pictures_DACH_ShapeLayout_RelativeWhiteSpaceAverage}). This encompasses the development of a new algorithm and parameter tuning which  makes it subject to future work.\\
Further, an analysis on specific graph properties and our shape change enhancements is for sure valuable in the future. This would provide in in-depth characterization for which DAGs our layout is particularly suitable, where performance drops need to be expected, and even if it would be advisable to invent further criteria for specific cases. This would be a suitable guidance for a future user of our layout. He would be spared of time-consuming try an error attempts. As it is the case for every layout - no layout is equally suitable for all graph types and properties.

    \section*{Acknowledgements}

We greatly benefited from the feedback of Prof. G\"{u}nther Wallner. We would like to thank you very much for this.

\bibliographystyle{abbrvurl}
\bibliography{meine_Diss_references}
\end{document}